\documentclass{aa}  
\usepackage{graphicx}
\usepackage{float} 
\usepackage[varg]{txfonts}
\usepackage{booktabs}
\usepackage{siunitx}
\usepackage{capt-of}

\bibpunct{(}{)}{;}{a}{}{,}
\usepackage{xcolor}
\usepackage[colorlinks=true,linkcolor=blue,citecolor=blue,urlcolor=blue]{hyperref}

\begin{document}

   \title{The role of radiative torques in the molecular cloud core L43}

   \author{M. L. Scheiter
          \and
          S. Wolf
          }

   \institute{University of Kiel, Institute of Theoretical Physics and Astrophysics,
                Leibnizstrasse 15, 24118 Kiel, Germany\\
              \email{[mscheiter;wolf]@astrophysik.uni-kiel.de}
             }

   \date{Received month day, year; accepted month day, year}
 
  \abstract
    {Polarized emission from interstellar dust grains is commonly used to infer information about the underlying magnetic field from the diffuse interstellar medium to molecular cloud cores. Therefore, the ability to accurately determine properties of the magnetic field requires a thorough understanding of the dust alignment mechanism.} 
    {We investigate the influence of anisotropic radiation fields on the alignment of dust particles by magnetic fields, known as radiative torque (RAT) alignment. Specifically, we take advantage of the unique spatial configuration of the molecular cloud core L43, which contains an embedded yet optically visible star acting as a local source of anisotropic illumination.} 
    {Based on polarization maps obtained at wavelengths of $154\,\mu\mathrm{m}$ (SOFIA/HAWC+), as well as $450\,\mu\mathrm{m}$ and $850\,\mu\mathrm{m}$ (JCMT/SCUBA-2), which show variations in the degree and angle of polarized emission across all wavelengths, we applied the differential measure analysis method to infer magnetic field strengths and analyze the global polarization spectrum of this source.} 
    {We derived plane-of-sky magnetic field strengths ranging from approximately 13 to 60 $\mu\mathrm{G}$, varying with wavelength, and find a negative slope of the polarization spectrum. Compared to 3D radiative transfer simulations, this finding can be attributed, at least partially, to variations in dust properties and temperatures along the line of sight. However, the additional influence of variations in the magnetic field orientation along the line of sight cannot be ruled out.} 
    {Our results favor radiative torques as the primary alignment mechanism, as they indicate that the degree of polarization is dependent on temperature and hence the strength of the local radiation field.}


   \maketitle

\section{Introduction}
Magnetic fields are thought to influence the process of star formation by regulating the rate at which dense interstellar material collapses under its own gravity (see, e.g., \citealp{krumholz_role_2019} and \citealp{pattle_magnetic_2022} for reviews). One method that has proven useful in inferring the morphology and strength of magnetic fields in star-forming regions is based on polarization measurements. As first pointed out by \citet{hall_observations_1949} and \citet{hiltner_polarization_1949}, the optical to near-infrared polarization of stellar radiation arises from magnetically aligned interstellar dust grains. The mechanisms responsible for the alignment continue to be an active topic of research today. Early explanations, such as alignment through paramagnetic dissipation \citep{davis_polarization_1951}, have struggled to hold up against observational testing. Today, the most promising candidate for a theory of dust grain alignment is radiative torque (RAT) alignment (\citealp{dolginov_orientation_1976,draine_radiative_1996,lazarian_radiative_2007,andersson_interstellar_2015}). Here, differences in the scattering efficiency of left- and right-circularly polarized light of an anisotropic radiation field cause dust grains with net helicity to spin up to suprathermal rates. If the material of the dust grain is paramagnetic, this rotation then leads to the magnetization of the grain through the Barnett effect \citep{barnett_magnetization_1915} and consequently causes the grain to align preferentially with its short axis parallel to the magnetic field lines. At far-infrared to submillimeter wavelengths, this preferred orientation causes a net polarization of the thermally emitted light that is perpendicular to the plane-of-sky (POS) component of the magnetic field. To accurately infer the strength and morphology of the underlying magnetic field, it is therefore necessary to first constrain the dust properties, as well as the efficiency of grain alignment across different environments.\\
In this work, we study the far-infrared polarization spectrum \citep{hildebrand_far-infrared_1999} -- namely, variations in the degree of linear polarization with wavelength. In particular, we study the molecular cloud core L43, as it contains a dense starless core in close proximity to a young stellar object (YSO), acting as an anisotropic radiation source. Previous work (e.g., \citealp{santos_far-infrared_2019,michail_far-infrared_2021,seifried_origin_2023,tram_understanding_2024}) indicates that changes in the polarization spectrum are driven by temperature differences between dust species that align with the magnetic field (e.g., silicate grains) and those that are thought not to align (e.g., carbonaceous grains), as well as differences in the local temperatures along the line of sight (LOS). The latter effect is also termed the "heterogeneous cloud effect" (HCE; \citealp{michail_far-infrared_2021}). As local temperature differences are inherently coupled to the strength of the radiation field that heats the grains, these differences cause the efficiency of the RAT mechanism to vary as well. As polarization measurements are constrained to the POS component of the magnetic field, changes in the orientation of the field lines influence the total degree of polarization. Therefore, by measuring the polarization of thermal reemission radiation at multiple wavelengths, with each preferentially tracing grains at different temperatures and depths, variable orientations of the magnetic field along the LOS also influence the shape of the polarization spectrum. We have attempted to separate the effect of magnetic field orientation from HCE by comparing observations with corresponding simulations of a simple molecular cloud core, modeled with the 3D Monte Carlo radiative transfer code POLARIS \citep{reissl_radiative_2016}.\\
This paper is structured as follows: In Sect.~\ref{L43Kap}, we give a brief description of the source, L43. Sect.~\ref{ObsDataKap} describes the observations, as well as the data reduction and preparation techniques used. In Sect.~\ref{ObsResultsKap}, we analyze the polarization data and explain how we derived temperature and column density maps, which we used to calculate magnetic field strengths. In Sect.~\ref{PolarisKap}, we explain how we employed the radiative transfer code POLARIS to construct a simple cloud core model for further analysis of the spatially resolved polarization spectrum. Finally, we give a brief discussion of our results in Sect.~\ref{DiscussionKap} and summarize our conclusions in Sect.~\ref{ConclusionsKap}.

\section{Source description}\label{L43Kap}
L43 is a dark filamentary cloud with a Lynds opacity class of six (the highest opacity class in the Lynds catalog \citealp{lynds_catalogue_1962}). It contains a dense starless core to the east and a partially embedded Class I protostar (classified by \citealp{chen_spitzer_2009}), commonly referred to as RNO 91 ("red nebulous object"; \citealp{cohen_red_1980}), to the west. The cloud lies in Ophiuchus North (Oph N), the northern part of the Ophiuchus star-forming region (see, e.g., \citealp{hatchell_spitzer_2012,alves_hp2_2025} for an overview of the region). We assume that the distance to L43 is not significantly different from the rest of the cloud complex, for which \citet{de_geus_physical_1989} find $d = 125 \pm 25~\mathrm{pc}$.\\
The protostar RNO 91 is associated with a bipolar CO outflow \citep{bence_l_1998,lee_co_2002,lee_outflow_2005}, seen in $^{12}\mathrm{CO}$ $J=1-0$, $J=2-1$ and $J=3-2$ spectral line data. By comparing the $^{12}\mathrm{CO}$ data with I-band continuum data, \citet{mathieu_l43_1988} pointed out that the southern lobe of the outflow has influenced the dust morphology of the cloud by clearing a U-shaped cavity around RNO 91. This influence was also noticed by \citet{ward-thompson_first_2000}, who studied the polarized flux of L43 with the SCUBA/SCUPOL polarimeter at the James Clerk Maxwell Telescope (JCMT) and noted that the outflow has likely influenced the morphology of the underlying magnetic field. More recent observations with its successor SCUBA-2/POL-2 by \citet{karoly_jcmt_2023} support this as well. This possible interaction between the outflow and the magnetic field complicates the determination of magnetic field strengths through polarization measurements from thermal dust emission in this region. Nonetheless, \citet{crutcher_scuba_2004} apply the Davis-Chandrasekhar-Fermi (DCF) method (\citealp{davis_strength_1951,chandrasekhar_problems_1953}) to the densest region of L43, assuming this region is still unaffected by the outflow. They find POS field strengths of $\mathrm{B}_{\mathrm{POS}} \approx 160~\mu\mathrm{G}$ for an area with an average volume density of $\mathrm{n}_\mathrm{H_2}\approx3.8\times10^5\mathrm{cm}^{-3}$. \citet{karoly_jcmt_2023} divide L43 in three different subregions and find POS field strengths between $\mathrm{B}_{\mathrm{POS}}\approx (40-257)\,\mu\mathrm{G}$ and volume densities of $\mathrm{n}_\mathrm{H_2}\approx(0.6-4.9)\times10^5\mathrm{cm}^{-3}$.\\

\section{Observations and data reduction}\label{ObsDataKap}
A summary of all the instruments used, along with their measuring wavelengths, pixel, and beam sizes, is given in Table~\ref{InstrumentsTab}.\\
\subsection{SOFIA/HAWC+}
SOFIA/HAWC+ \citep{harper_hawc_2018} D-band observations ($154\,\mu\mathrm{m}$) of L43 were carried out on 11 June 2022 as part of observation cycle 9 (Proposal: 09\_0159, Obs. ID: P\_2022-06-11\_HA\_F886-09\_0159\_4-044), utilizing the on-the-fly mapping (OTFMAP) mode\footnote{\url{https://irsa.ipac.caltech.edu/data/SOFIA/docs/instruments/handbooks/HAWC\_Handbook\_for\_Archive\_Users\_Ver1.0.pdf}}. The data were processed using the "HAWC DRP v3.0.0" pipeline\footnote{\url{https://irsa.ipac.caltech.edu/data/SOFIA/docs/data/data-pipelines/}} (see, e.g., \citealp{santos_far-infrared_2019}), resulting in "Level 4" (science quality) data. These include images of the total intensity (Stokes $I$), the linearly polarized intensities (Stokes $Q$ and Stokes $U$), as well as the degree of polarization $p^{\prime}$, the debiased degree of polarization $p$ (see Sect.~\ref{DataPrepKap}), the polarization angle $\theta$, and their corresponding errors.\\

\subsection{SCUBA-2/POL-2}\label{SCUBADataKap}
L43 was observed at $450~\mu\mathrm{m}$ and $850~\mu\mathrm{m}$ simultaneously with the Submillimetre Common-User Bolometer Array 2 (SCUBA-2; \citealp{holland_scuba-2_2013}) and its polarimeter POL-2 on the James Clerk Maxwell Telescope (JCMT) as part of the B fields In STar Forming Regions Observations-3 (BISTRO-3) large survey program (\citealp{ward-thompson_first_2017}, Project ID: M20AL018). Twenty-seven repeated observations with an integration time of 30 minutes each were carried out in the time frame between February 2020 and March 2021. One observation is flagged as questionable and omitted from the data reduction process for both wavelengths. Additionally, data from one observation at $450~\mu\mathrm{m}$ are incomplete and were also omitted. For the data reduction, we used the Sub-Millimetre User Reduction Facility (SMURF\footnote{\url{http://www.starlink.ac.uk/docs/sun258.pdf}}; \citealp{chapin_scuba-2_2013}), which is part of the Starlink software package \citep{currie_starlink_2014}. SMURF contains an iterative reduction routine for POL-2 data, called pol2map, which is used to clear the raw bolometer time streams from atmospheric and instrumental noise and create separate maps of Stokes $I$,$Q$, $U$, and their corresponding errors. We closely follow the reduction steps laid out by \citet{karoly_jcmt_2023}, resulting in one reduction with a pixel size of $4^{\prime\prime}$ that is used in the calculation of column density and temperature maps (see Sect.~\ref{BbfitsKap}) and a second reduction with a pixel size of $8^{\prime\prime}$ that is used to create a polarization vector catalog, which itself is binned to $12^{\prime\prime}$.\\ 
The reduced flux data from SCUBA-2, given in units of $\mathrm{pW}$, are converted to $\mathrm{Jy\,beam^{-1}}$ by applying a flux conversion factor (FCF), calculated from calibration sources with well-known fluxes. We apply FCFs of $(472\pm 76)\,\mathrm{Jy\,beam^{-1}\,pW^{-1}}$ and $(495\pm 32)\,\mathrm{Jy\,beam^{-1}\,pW^{-1}}$ multiplied by additional factors of $1.96$ and $1.35$ for $450\,\mu\mathrm{m}$ and $850\,\mu\mathrm{m}$, respectively \citep{mairs_decade_2021}. The latter factors are additional correction factors applied to cover flux losses induced by POL-2.\\
The $^{12}\mathrm{CO}\;(3-2)$ molecular line emission is known to contaminate the $850\,\mu\mathrm{m}$ data obtained with SCUBA-2 \citep{drabek_molecular_2012}, leading to an overestimation of the continuum flux. To correct for this overestimation, we follow the instructions on the East Asian Observatory website\footnote{\url{https://www.eaobservatory.org/jcmt/science/reductionanalysis-tutorials/scuba-2-dr-tutorial-5/}}, by introducing $^{12}\mathrm{CO}\;(3-2)$ line emission data from the HARP/ACSIS instrument \citep{buckle_harpacsis_2009} (Proposal ID: M07AU11) as a negative fakemap\footnote{\url{http://www.starlink.ac.uk/docs/sun258.pdf}} to the reduction with pol2map. The percentage flux originating from CO contamination amounts to $10\%-15\%$ to the west of RNO 91 and less than $10\%$ to the east.

\subsection{Herschel PACS and SPIRE}\label{HerschelDataKap}
In order to calculate column density and temperature maps (see Sect.~\ref{BbfitsKap}), we obtained data in the 160~$\mu\mathrm{m}$ band of the Photometer Array Camera and Spectrometer (PACS; \citealp{poglitsch_photodetector_2010}) and in the 250~$\mu\mathrm{m}$, 350~$\mu\mathrm{m}$ and 500~$\mu\mathrm{m}$ band of the Spectral and Photometric Imaging Receiver (SPIRE; \citealp{griffin_herschel-spire_2010}) Instruments on board the Herschel Space Observatory \citep{pilbratt_herschel_2010}, which are publicly available at the Herschel Science Archive\footnote{\url{https://archives.esac.esa.int/hsa/whsa/}}. The Herschel data (Obs. ID: OT1\_jhatchel\_1) were acquired in the time frame between 17 February 2013 and 17 March 2013, utilizing the PACS/SPIRE Parallel Mode. Observations made in the PACS/SPIRE Parallel Mode are processed through the SPIRE Photometer Large Map pipeline and stored as a FITS file. In contrast to the reduction of the PACS/SPIRE data, the SCUBA-2 reduction pipeline needs to account for atmospheric noise. This leads to differences in the amount of recovered large-scale emission between the instruments. To account for this discrepancy, we follow the method described by \citet{sadavoy_herschel_2013}, where the Herschel flux maps are inserted into the SCUBA-2 data reduction pipeline as a fakemap. The resulting flux maps are sensitive to the same emission as the SCUBA-2 data. We adopt flux calibration uncertainties of 20\% for the PACS data and 10\% for the SPIRE data \citep{chuss_hawcsofia_2019}.\\

\subsection{Nobeyama Radio Observatory/FOREST}\label{NROKap}
The calculation of magnetic field strengths (see Sect.~\ref{BfieldKap}) requires information about the turbulent gas kinematics of L43. For this purpose, we made use of four N$_2$H$^+$ ($1-0$) line emission observations taken with the superconductor-insulator-superconductor receiver FOREST \citep{Minamidani_2016} at the 45-m telescope operated by the Nobeyama Radio Observatory (NRO). The observations (Obs. ID's: 20180504020849, 20180505015815, 20180506020433, 20190401033107) were taken between 4 May 2018 and 1 April 2019 and are publicly available at the Nobeyama-45m / ASTE Science Data Archive\footnote{\url{https://nobeyama-archive.nao.ac.jp/user/index.html}}. The data is stored in position-position-velocity cubes with a spectral resolution of 0.098 $\mathrm{km\,s^{-1}}$. We combined the observations utilizing the reprojection package of Spectral-Cube\footnote{\url{https://spectral-cube.readthedocs.io/en/latest/reprojection.html}}. The resulting spectra have an average rms noise of $\sim 0.5~\mathrm{K}$, which allows us to only consider data exceeding an S/N-threshold of 3.

\begin{table}
\caption{Observing wavelengths as well as beam and pixel sizes of the individual instruments.}
\label{InstrumentsTab}
\centering
\begin{tabular}{c c c c}
\hline\hline
Instrument & Wavelength & Beam size & Pixel size \\
 & [$\mu\mathrm{m}$] & [$^{\prime\prime}$]& [$^{\prime\prime}$] \\
\hline
   SOFIA/HAWC+$^{(a)}$ & 154 & 14.0 & 3.4 \\
   Herschel/PACS$^{(b)}$ & 160 & 11.3 & 3.2 \\
   Herschel/SPIRE$^{(c)}$ & 250 & 18.1 & 3.0 \\
   Herschel/SPIRE$^{(c)}$ & 350 & 25.2 & 5.0 \\
   Herschel/SPIRE$^{(c)}$ & 500 & 36.6 & 7.0 \\ 
   SCUBA-2/POL-2$^{(d)}$ & 850 & 14.6 & 4.0 \\
   NRO/FOREST$^{(e)}$ & 3217 & 22.5 & 6.9 \\
\hline          
\end{tabular}
\tablefoot{\\
$^{(a)}$ \citet{harper_hawc_2018}\\
$^{(b)}$ \citet{poglitsch_photodetector_2010}\\
$^{(c)}$ \citet{griffin_herschel-spire_2010}\\
$^{(d)}$ \citet{dempsey_scuba-2_2013}\\
$^{(e)}$ \citet{Minamidani_2016}}
\end{table}

\subsection{Data preparation}\label{DataPrepKap}
\noindent For an accurate comparison of the flux and polarization maps obtained with the different instruments, their pixel- and beam sizes are standardized. For the polarization and spectral line data, the command WCSALIGN\footnote{\url{https://starlink.eao.hawaii.edu/docs/sun95.htx/sun95ss203.html}} from Starlink's KAPPA package is applied to the SOFIA/HAWC+ and NRO data. This command takes the SCUBA-2/POL-2 observations as a reference, aligns the World Coordinate System (WCS) and adjusts the pixel size of the data to match the pixel size of the vector catalog ($12^{\prime\prime}$). The pixel sizes used for the column density and temperature fits (see Sect.~\ref{BbfitsKap}) are standardized to $4^{\prime\prime}$, by applying the adaptive anti-aliasing algorithm reproject\_adaptive\footnote{\url{https://reproject.readthedocs.io/en/stable/}} to the Herschel PACS and SPIRE data.\\
To standardize the beam sizes, we follow \citet{chuss_hawcsofia_2019} and \citet{michail_far-infrared_2021}, who smooth the images with a smaller beam size with a Gaussian kernel. The FWHM of these smoothing kernels are determined via
\begin{equation}\label{smoothingEq}
    \mathrm{FWHM_{smooth}} = \sqrt{\mathrm{FWHM^2_{ref}} - \mathrm{FWHM^2_{\lambda}}},
\end{equation}
where $\mathrm{FWHM_{ref}}$ describes the FWHM of the reference beam (the largest FWHM in the dataset) and $\mathrm{FWHM_{\lambda}}$ the original FWHM of the instrument beam at wavelength $\lambda$.\\
We apply this correction to the intensity data and the polarization data independently. The intensity data are smoothed to a resolution of $36.6^{\prime\prime}$, corresponding to the Herschel/SPIRE beam size at $500~\mu\mathrm{m}$, while the polarization data are smoothed to a resolution of $14.6^{\prime\prime}$, corresponding to the effective beam size of SCUBA-2/POL-2 at $850~\mu\mathrm{m}$ (see Table~\ref{InstrumentsTab}).\\
\noindent Unlike the SOFIA/HAWC+ data, the degree of linear polarization $p^{\prime}$ and angle $\theta$ are not precalculated for the SCUBA-2/POL-2 data. These values are determined from the Stokes $I$, $Q$ and $U$ data and calculated from the equations
\begin{equation}
    p^{\prime} = \frac{1}{I}\sqrt{Q^{2}+U^{2}}, \qquad \theta = \frac{1}{2}\tan^{-1}\left(\frac{U}{Q}\right).
\end{equation}

\noindent Since the degree of polarization is a positive-definite quantity (see, e.g., \citealp{vaillancourt_placing_2006}), we attempt to correct for this bias by following \citet{wardle_linear_1974} in introducing the debiased polarization fraction $p$, calculated from the Stokes parameters $I$, $Q$ and $U$ and their respective uncertainties $\delta Q$ and $\delta U$
\begin{equation}
    p = \sqrt{p^{\prime}-\delta p^{\prime}} = \frac{1}{I}\sqrt{Q^{2}+U^{2}-\frac{Q^{2}\delta Q^{2} + U^{2}\delta U^{2}}{Q^{2}+U^{2}}},
\end{equation}
where $\delta p^{\prime}$ is the corresponding uncertainty of $p^{\prime}$. Propagating the uncertainty of $p$ yields
\begin{equation}
    \delta p = \sqrt{\frac{Q^{2}\delta Q^{2}+U^{2}\delta U^{2}}{I^{2}(Q^{2}+U^{2})}+\frac{\delta I^{2}(Q^{2}+U^{2})}{I^{4}}},
\end{equation}
where $\delta I$ describes the uncertainty for the total intensity $I$. For the uncertainty on $\theta$ follows
\begin{equation}
    \delta \theta = \frac{1}{2}\frac{\sqrt{Q^{2}\delta U^{2} + U^{2}\delta Q^{2}}}{Q^{2}+U^{2}}\times \frac{180^\circ}{\pi}.
\end{equation}
The following criteria are applied in the selection of the polarization vectors: 
\begin{equation}\label{CutoffEq}
    \frac{I}{\delta I}> 15, \qquad \frac{p}{\delta p} > 2.
\end{equation}

\section{Observational results}\label{ObsResultsKap}
\begin{figure*}
  \centering
  \includegraphics[width=\textwidth]{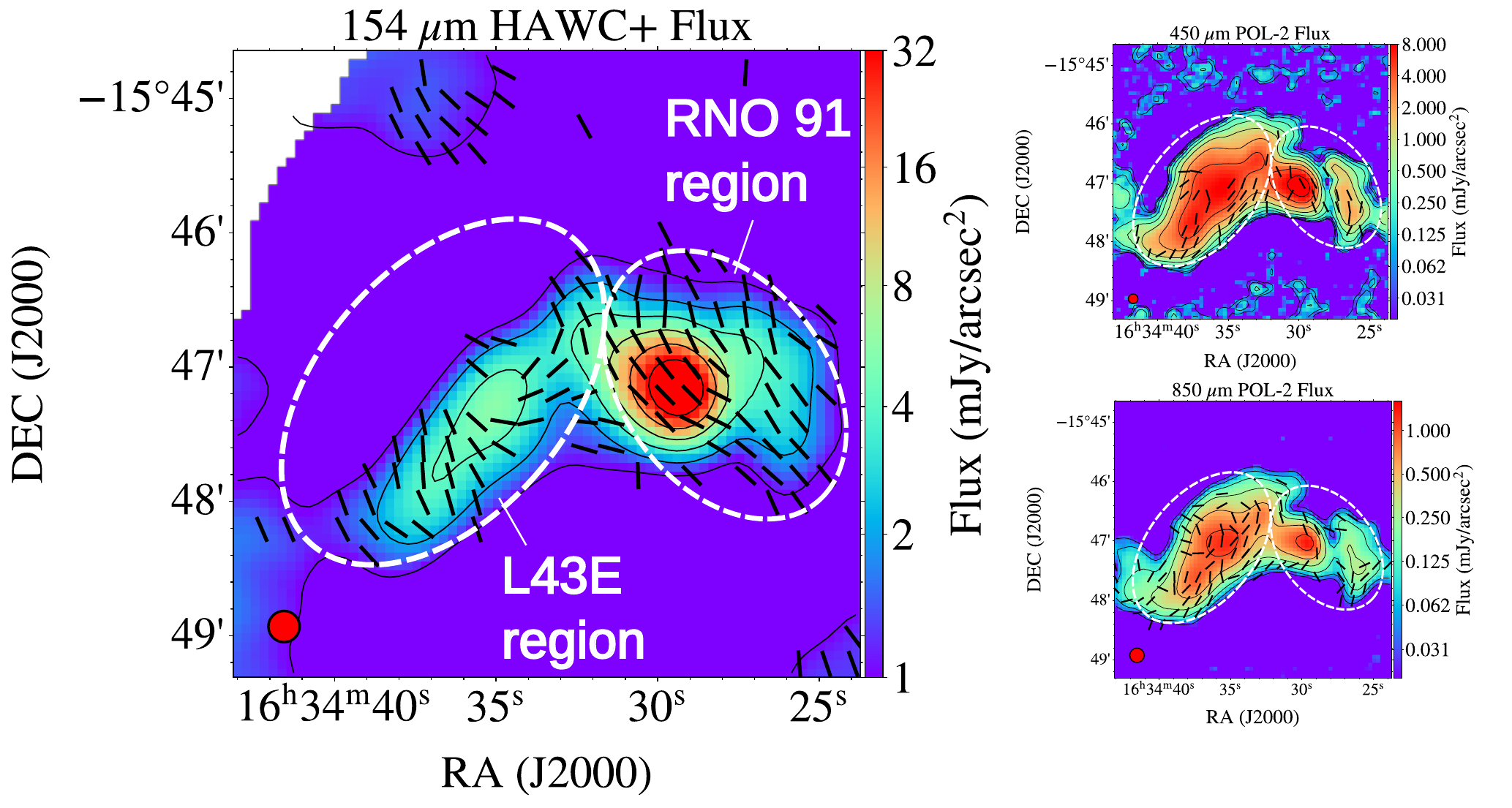}
  \caption{Dust continuum maps of the L43 region. Black lines indicate the polarization vectors. The vectors are rotated by $90^{\circ}$ to indicate the magnetic field direction and are binned to the same $12^{\prime\prime}$ grid. Black contour lines mark regions where the flux has doubled. White dashed ellipses show the two regions of interest that are referred to throughout this work. The effective beam size of each instrument is indicated with a red circle. Left: $154\,\mu\mathrm{m}$ continuum emission (SOFIA/HAWC+). Top right: $450\,\mu\mathrm{m}$ continuum emission (SCUBA-2/POL-2). Bottom right: $850\,\mu\mathrm{m}$ continuum emission (SCUBA-2/POL-2).}
  \label{FluxMapFig}
\end{figure*}

\subsection{Emission structure}
Fig.~\ref{FluxMapFig} shows the dust continuum emission at $154\,\mu\mathrm{m}$ (left), $450\,\mu\mathrm{m}$ (top right), and $850\,\mu\mathrm{m}$ (bottom right), along with the rotated polarization vectors that fall above the S/N-cutoff defined in Eq.~\ref{CutoffEq}. The impact of the irradiation by the YSO RNO 91 is clearly seen at shorter wavelengths, whereas for the $850\,\mu\mathrm{m}$ observation, the flux decreases to a similar peak level as in L43E. These differences in the emission structure are likely to be attributed to variations in the local dust temperature. Note that the SOFIA/HAWC+ observation does not recover the extended emission seen in the northeastern part of the SCUBA-2/POL-2 observations in the L43E region.\\

\subsection{Polarization properties}\label{polpropskap}
\subsubsection{Magnetic field morphology}
The magnetic field structure inferred from the polarization vectors in Fig.~\ref{FluxMapFig} shows a clear wavelength dependence. As \citet{karoly_jcmt_2023} points out, the inferred magnetic field direction in the southern region of L43E coincides well with the cavity walls created by the molecular outflow associated with the YSO, suggesting that the magnetic field has been influenced by the outflow. This influence is also seen in the $450\,\mu\mathrm{m}$ emission, but is absent in the $154\,\mu\mathrm{m}$ emission. Fig.~\ref{PThetaHistFig} shows the distribution of the degree of polarization, as well as the distribution of the rotated polarization angles. The estimated outflow direction of $150^\circ \pm 10^\circ$ \citep{karoly_jcmt_2023} coincides well with the circular mean of the $450\,\mu\mathrm{m}$ and $850\,\mu\mathrm{m}$ data in L43E, which are $151^\circ$ and $153^\circ$, respectively. The stark difference in magnetic field morphology as well as emission structure between the SOFIA/HAWC+ data and the SCUBA-2/POL-2 data suggests that we might be observing different layers of the cloud, caused by differences in temperatures along the LOS. While the $154\,\,\mu\mathrm{m}$ emission traces the warmer, diffuse regions of L43, the $450\,\mu\mathrm{m}$ and $850\,\mu\mathrm{m}$ emission trace the inner, colder regions.\\

\subsubsection{The polarization hole in L43}\label{PolHoleKap}
\begin{figure}[htbp]
  \centering
  \includegraphics[width=\columnwidth]{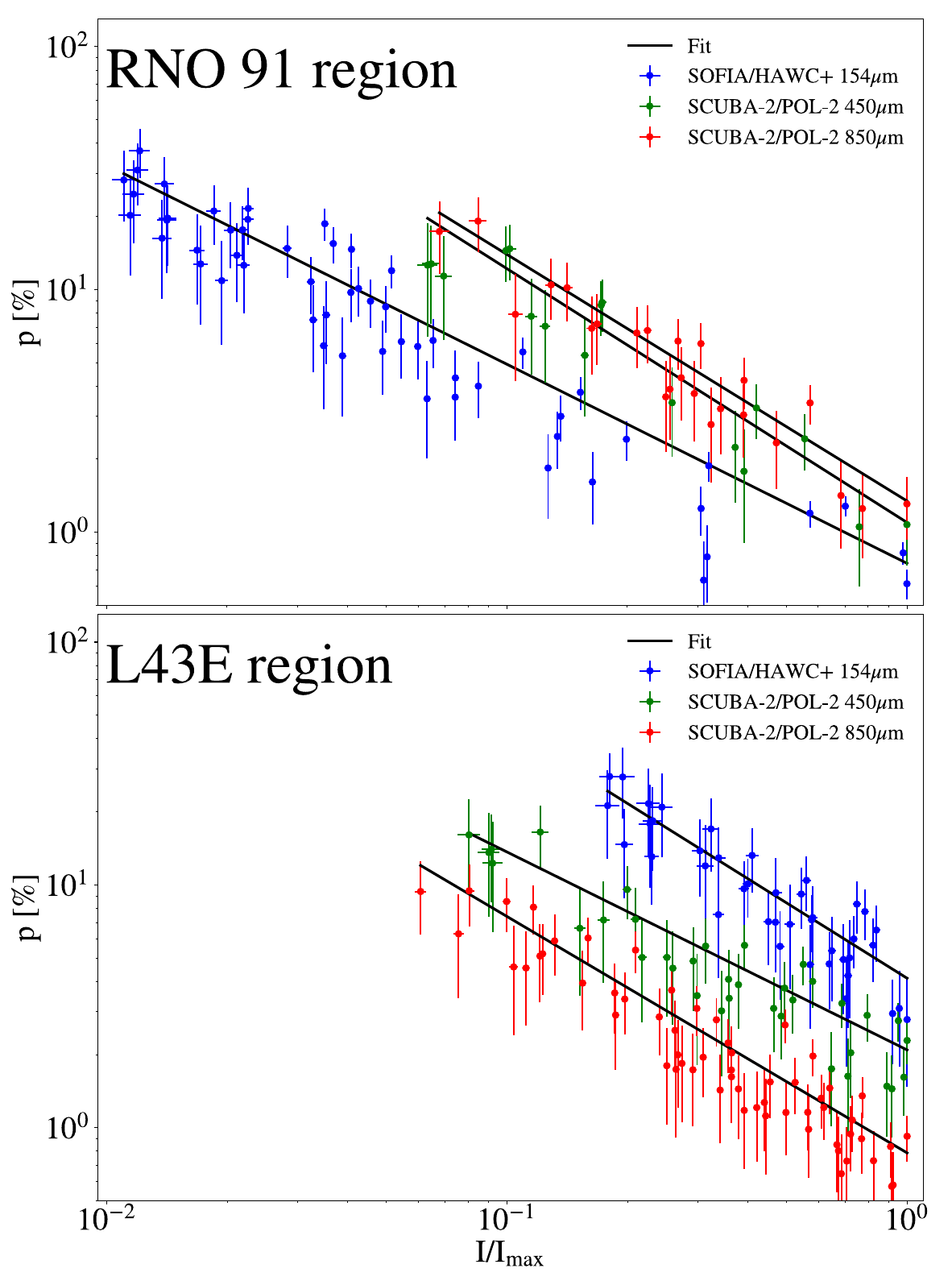}
  \caption{Degree of polarization as a function of the normalized intensity in the RNO 91 region (top panel) and the L43E region (bottom panel). The black lines indicate the best fit of Eq.~\ref{PvsIEq} to each dataset. In the L43E region, we find $a_2 = 1.03^{0.10}_{-0.10}$ ($\chi^2_{\mathrm{red}}=0.63$) at $154\,\mu\mathrm{m}$, $a_2 = 0.81^{0.08}_{-0.07}$ ($\chi^2_{\mathrm{red}}=0.73$) at $450\,\mu\mathrm{m}$ and $a_2 = 0.97^{0.05}_{-0.05}$ ($\chi^2_{\mathrm{red}}=0.99$) at $850\,\mu\mathrm{m}$. In the RNO 91 region, we find $a_2 = 0.82^{0.02}_{-0.02}$ ($\chi^2_{\mathrm{red}}=1.79$) at $154\,\mu\mathrm{m}$, $a_2 = 1.04^{0.10}_{-0.10}$ ($\chi^2_{\mathrm{red}}=0.73$) at $450\,\mu\mathrm{m}$ and $a_2 = 1.02^{0.09}_{-0.09}$ ($\chi^2_{\mathrm{red}}=0.69$) at $850\,\mu\mathrm{m}$.}
  \label{PvsIFig}
\end{figure}

Previous analyses of near-infrared polarization properties in molecular clouds have shown that there exists a correlation between polarization efficiency, denoted as the degree of polarization due to extinction, per optical depth $p_{\mathrm{ext},\lambda}/\tau_\lambda$, with visual extinction $A_V$ \citep{whittet_efficiency_2008}. This correlation, often referred to as the term "polarization hole", is usually parameterized in the form of a power law
\begin{equation}
    p_{\mathrm{ext},\lambda}/\tau_\lambda = a_1 A_V^{-a_2}.
\end{equation}
Here, $a_2 \approx 0$ indicates that alignment is retained throughout the denser regions of a cloud, while $a_2 \approx 1$ is interpreted as polarized emission originating only from the outermost layer of dust \citep{andersson_interstellar_2015}. However, additional effects, such as a decrease in polarization due to changes in the direction of the magnetic field lines on scales smaller than the beam size or the non-Gaussian nature of the noise on $p$ \citep{vaillancourt_placing_2006,pattle_jcmt_2019} potentially lead to an overestimation of $a_2$.\\
In the context of far-infrared and submillimeter polarimetry, it is common to use the total intensity $I$ as a proxy for the visual extinction $A_V$\footnote{See \citet{pattle_jcmt_2019} for a discussion of the validity of this proxy.}. Together with the relation $p_{\mathrm{ext},\lambda} = - \tau_\lambda p_{\mathrm{emit},\lambda}$\footnote{The minus sign is a convention used to indicate the $90^\circ$ flip in polarization angle \citep{lazarian_tracing_2007}.}\citep{andersson_interstellar_2015}, the dependency that is fitted to the polarization data becomes
\begin{equation}\label{PvsIEq}
    p_{\mathrm{emit},\lambda} = a_1 \left(\frac{I}{I_{\mathrm{max}}}\right)^{-a_2},
\end{equation}
where $I_{\mathrm{max}}$ denotes the intensity maximum in the considered region.\\
Fig.~\ref{PvsIFig} shows the fit function from Eq.~\ref{PvsIEq} applied to the data obtained at all three wavelengths in our regions of interest. Here, the fitting is done via the Markov chain Monte Carlo method. Every dataset is well fit\footnote{Under the assumption of Gaussian noise.}, indicated by values of $\chi^2_{\mathrm{red}} < 1$, with the only exception being the SOFIA/HAWC+ data in the RNO 91 region, where we find $\chi^2_{\mathrm{red}} = 1.79$. The exponent $a_2$ shows little variation between regions and wavelengths, with values ranging from $0.81 - 1.03$ in the L43E region and $0.82 - 1.04$ in the RNO 91 region. These findings indicate that the polarization efficiency in the dense regions of L43 might be heavily attenuated. To assess whether the positive bias in $p$ is still affecting our results, we repeat the fits with a stricter S/N-cutoff of $p/\delta p > 3$ and find no significant deviation from the previous values of $a_2$. One notable result is that we still see a strong depolarization occurring in the RNO 91 region, even though the presence of an embedded radiation source is expected to aid the alignment. To check whether this result still holds true in the direct vicinity of RNO 91, we repeat the fit in a circle with a $20^{\prime\prime}$ radius around the intensity peak of the RNO 91 region with the SOFIA/HAWC+ data and find $a_2 = 0.65^{0.06}_{-0.06}$. Due to the relatively low number of data points (14) in this region, the fit quality is slightly worse ($\chi^2_{\mathrm{red}} = 2.53$). 

\subsection{The far-infrared polarization spectrum of L43}\label{PolSpecKap}

\begin{figure*}
  \centering
  \includegraphics[width=\textwidth]{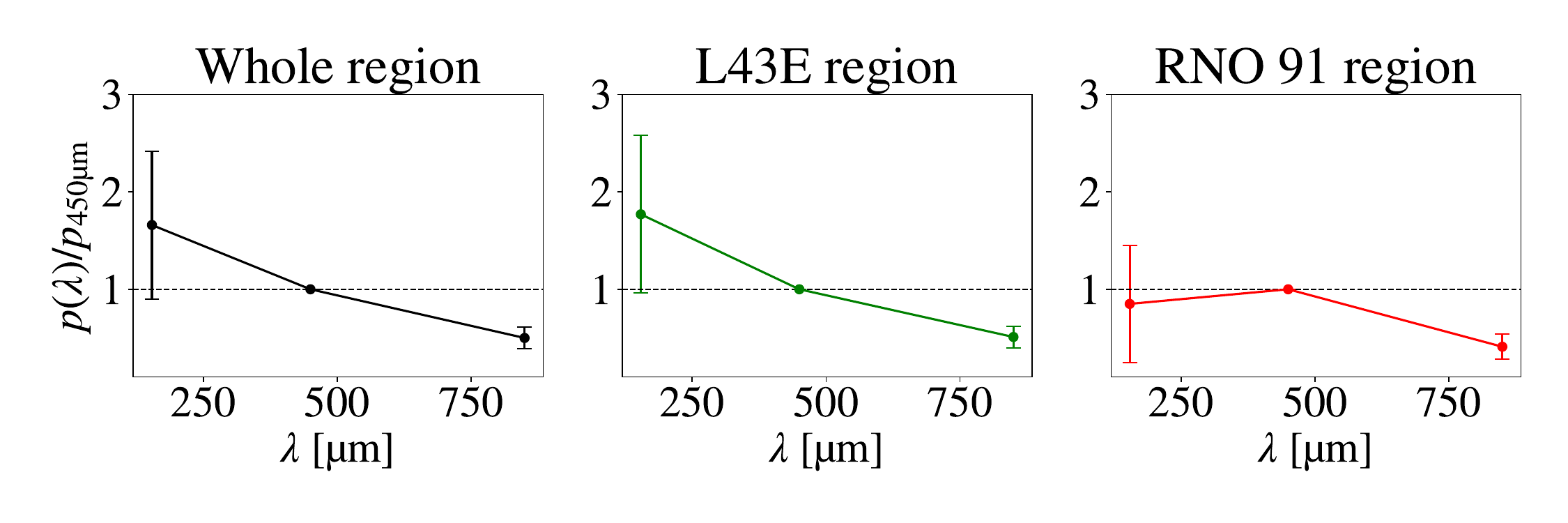}
  \caption{Global polarization spectra for the entire region (left panel), the L43E region (center panel), and the RNO 91 region (right panel). Points indicate the median, and error bars the MAD at a given wavelength. The spectra are normalized at $\lambda_0 = 450\,\mu\mathrm{m}$. A black dashed line indicates unity.}
  \label{PolSpecFig}
\end{figure*}

Given a singular dust species (e.g., silicate or carbonaceous grains), we do not expect variations in the degree of polarization at far-infrared wavelengths for any common dust species in the interstellar medium (ISM; \citealp{hildebrand_far-infrared_1999}). However, many previous observations have shown that such variations exist (e.g., \citealp{vaillancourt_submillimeter_2012,zeng_submillimeter_2013,santos_far-infrared_2019,michail_far-infrared_2021,cox_sofia_2025}). 
To construct the far-infrared polarization spectrum, we restrict our analysis to those data points, where measurements at all three wavelengths are available, that fulfill the criteria defined in Eq.~\ref{CutoffEq}. Previous studies have further constrained the selection of data points by limiting the analysis to points where the difference in polarization angles between wavelengths is below a certain threshold \citep{vaillancourt_analysis_2002}. This is done in an attempt to minimize the effects that a changing magnetic field along the LOS has on the polarization spectrum. Owing to the strong variability of polarization angles between wavelengths found in L43, we cannot apply this additional constraint and have to assume that the polarization spectrum is shaped in part by different inclinations of the magnetic field lines along the LOS.\\
Fig.~\ref{PolSpecFig} shows the global polarization spectrum for both considered subregions (as defined in Fig.~\ref{FluxMapFig}), as well as for the entire region. Data points indicate the median degree of polarization and error bars the median absolute deviation (MAD, see also Table~\ref{PolSpecTab}). Adhering to common practice, we normalize the spectra to the wavelength closest to $350\,\mu\mathrm{m}$, which in our case is $450\,\mu\mathrm{m}$. We find a decreasing spectrum in both subregions, with the only increase observed between $154\,\mu\mathrm{m}$ and $450\,\mu\mathrm{m}$ in the RNO 91 region. As \citet{seifried_origin_2023} mention in their semi-analytical modeling for a homogeneous cloud consisting of silicate and carbonaceous grains, we expect a monotonically increasing spectrum caused by higher average temperatures of carbonaceous grains (compared to silicate grains) that are believed not to align with the external magnetic field. When a given LOS contains two or more dust phases with different temperatures and alignment properties, the resulting spectrum is expected to become V-shaped, with a local minimum at around $350\,\mu\mathrm{m}$. The fact that our observed spectral slope is still negative between $450\,\mu\mathrm{m}$ and $850\,\mu\mathrm{m}$ implies that either the alignment of the dust in the cold phase is substantially impaired, or the inclination of the magnetic field is much higher in the cold phase (see Appendix~\ref{AppB} for details). Given the results obtained in Sect.~\ref{PolHoleKap}, we can see evidence of a reduced grain alignment efficiency in both subregions. However, a magnetic field influenced by a potentially heavily inclined outflow, for which \citet{lee_outflow_2005} find an inclination angle $i\approx (60 \pm 15)^\circ$ with respect to the POS, is likely contributing to the observed shape of the spectrum. Separating the contributions of these effects from observations alone is difficult. In Sect.~\ref{PolarisKap}, we investigate the effect of changing dust phases along the LOS.

\begin{table}
\caption{Median and MAD values of the global polarization spectrum in all considered regions.} 
\label{PolSpecTab}
\centering
\begin{tabular}{c c c c}
\hline\hline
\noalign{\vskip 2pt}
 & $\frac{p(154\,\mu\mathrm{m})}{p(450\,\mu\mathrm{m})}$ & $\frac{p(850\,\mu\mathrm{m})}{p(450\,\mu\mathrm{m})}$ & Nr. of points\\
\noalign{\vskip 2pt}
\hline
\noalign{\vskip 1pt}
   Entire region  & 1.66 $\pm$ 0.76 & 0.50 $\pm$ 0.11 & 21 \\
   L43E  & 1.77 $\pm$ 0.81& 0.51 $\pm$ 0.11 & 14 \\
   RNO 91  & 0.85 $\pm$ 0.60 & 0.41 $\pm$ 0.13 & 7\\ 
\hline 
\end{tabular}
\end{table}

\subsection{Column density and temperature maps}\label{BbfitsKap}
\begin{figure}[htbp]
  \centering
  \includegraphics[width=\columnwidth]{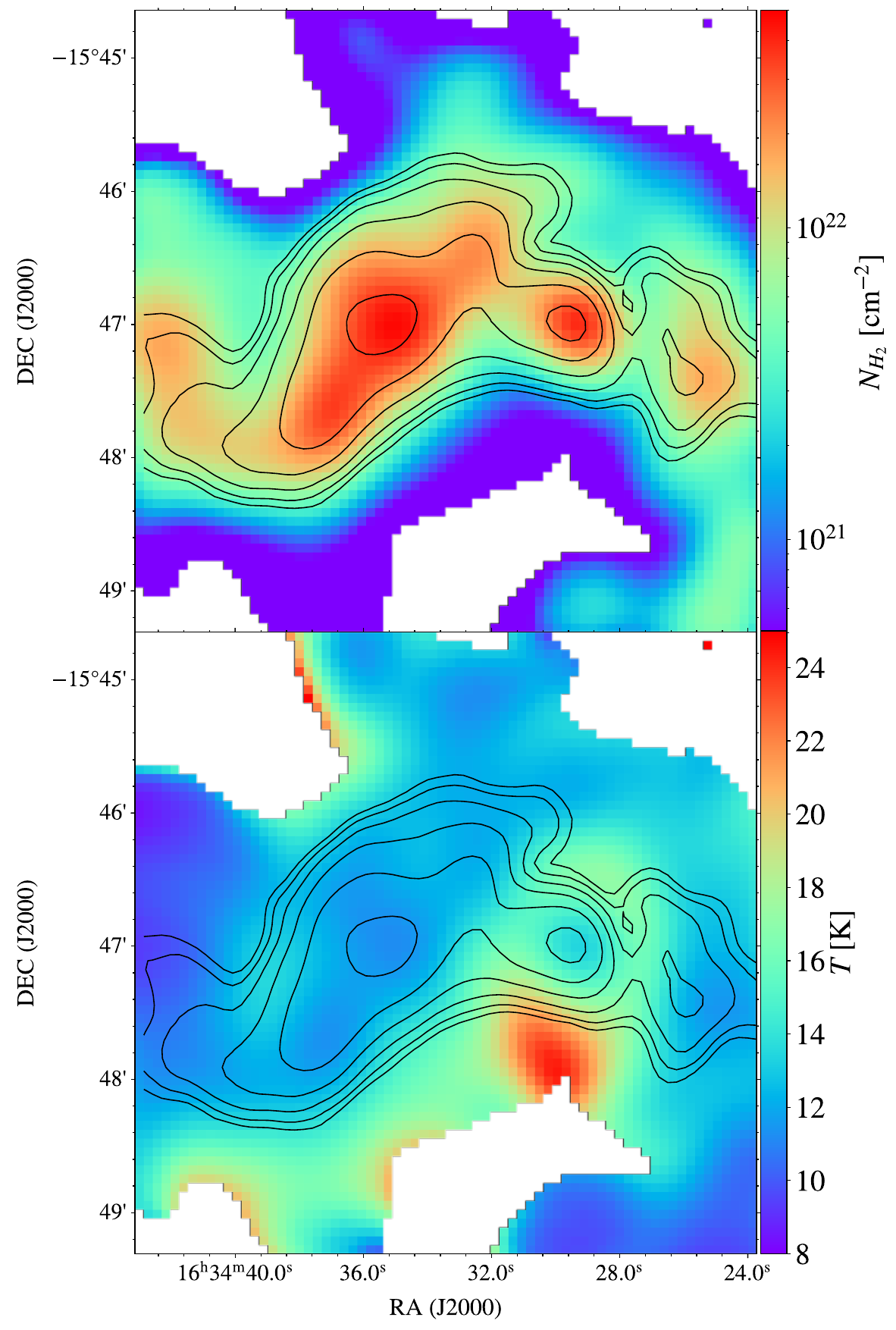}
  \caption{Spatially resolved column density (top panel) and LOS temperature (bottom panel) distributions, acquired by fitting Eq.~\ref{BbfitEq} to Herschel PACS/SPIRE and JCMT/SCUBA-2 data at observing wavelengths of $160\,\mu\mathrm{m}$, $250\,\mu\mathrm{m}$, $350\,\mu\mathrm{m}$, $500\,\mu\mathrm{m}$, and $850\,\mu\mathrm{m}$. Points where $\Delta_{T_\mathrm{d}}>5\,\mathrm{K}$ and $N_{\mathrm{H_2}}/\Delta_{N_{\mathrm{H_2}}}<1$ are excluded, where $\Delta_{T_\mathrm{d}}$ and $\Delta_{N_\mathrm{H_2}}$ refer to the errors on the respective fit parameter. Overlaid are contour lines of the $850\,\mu\mathrm{m}$ continuum emission from Fig.~\ref{FluxMapFig}.}
  \label{BBFitFig}
\end{figure}

Inferring the column density, LOS temperature, and optical depth of a molecular cloud requires a set of continuum measurements that sample the spectral energy distribution (SED) of the cloud. To do this, we use the measurements from the PACS and SPIRE instruments on board Herschel at $160\,\mu\mathrm{m}$, $250\,\mu\mathrm{m}$, $350\,\mu\mathrm{m}$, and $500\,\mu\mathrm{m}$ as described in Sect.~\ref{HerschelDataKap} in combination with the $850\,\mu\mathrm{m}$ measurement from SCUBA-2/POL-2 (Sect.~\ref{SCUBADataKap}). The beam and pixel sizes of all instruments were standardized according to Sect.~\ref{DataPrepKap} and the modified blackbody function \citep{hildebrand_determination_1983}
\begin{equation}\label{BbfitEq}
    I_\nu = (1-\exp{\left(\tau_\nu\right)})B_\nu(T_\mathrm{d})
\end{equation}
is fit to every pixel utilizing the Levenberg-Marquardt method. Here, $B_\nu(T_\mathrm{d})$ refers to the Planck function at dust temperature $T_\mathrm{d}$ and $\tau_\nu$ to the optical depth, described by the expression
\begin{equation}\label{OptDepthEq}
    \tau_\nu = \mu_{\mathrm{H}_2}m_\mathrm{H} N_{\mathrm{H}_2}\kappa_{\nu_0}\left(\frac{\nu}{\nu_0}\right)^\beta,
\end{equation}
where $\mu_{\mathrm{H}_2}$ is the mean molecular weight of molecular hydrogen, $m_\mathrm{H}$ the atomic mass of hydrogen, $N_{\mathrm{H}_2}$ the column density, $\kappa_{\nu_0}$ the dust opacity per unit mass for the reference frequency $\nu_0$ and $\beta$ the dust emissivity index. Keeping in line with other works (e.g.,$\,$\citealp{sadavoy_herschel_2013,karoly_jcmt_2023}), we adopt $\mu_{\mathrm{H}_2}=2.8$, $m_\mathrm{H}=1.00784\,\mathrm{u}$, $\kappa_{\nu_0} = 0.1\,\mathrm{cm^2g^{-1}}$, $\nu_0 = 10^{12}\,\mathrm{Hz}$ and $\beta = 1.8$. We simultaneously fit column density and temperature, as shown in Fig.~\ref{BBFitFig}. Pixels for which the error in $T_\mathrm{d}$ exceeds $5\,\mathrm{K}$, or the error in the column density is larger than the column density itself, are excluded.\\
For the L43E region, we find an average column density of $\overline{N_{\mathrm{H_2}}}= (1.38\pm0.34)\,10^{22}\,\mathrm{cm^{-2}}$ with a peak value of $N_{\mathrm{H_2,max}} = (4.72\pm1.25)\,10^{22}\,\mathrm{cm^{-2}}$ and a mean Temperature of $\overline{T_\mathrm{d}} = (12.5\pm0.4)\,\mathrm{K}$ with a peak value of $T_\mathrm{d,max} = (16.6\pm0.4)\,\mathrm{K}$. In the RNO 91 region, we find $\overline{N_{\mathrm{H_2}}}= (0.86\pm0.27)\,10^{22}\,\mathrm{cm^{-2}}$ and $N_{\mathrm{H_2,max}} = (3.83\pm1.55)\,10^{22}\,\mathrm{cm^{-2}}$ with temperatures $\overline{T_\mathrm{d}} = (15.2\pm0.9)\,\mathrm{K}$ and $T_\mathrm{d,max} = (23.5\pm3.9)\,\mathrm{K}$. The optical depth can now be calculated via Eq.~\ref{OptDepthEq}. For our shortest wavelength considered in this work, at $154\,\mu\mathrm{m}$, we find $\tau_{154\mu\mathrm{m},\mathrm{max}}\approx 0.03$.

\subsection{Magnetic field strength}\label{BfieldKap}
\subsubsection{Differential measure analysis}\label{DMAKap}
We apply the recently developed differential measure analysis (DMA; \citealp{Lazarian_2022}) technique to our polarization and spectral line data to infer magnetic field strengths for the two considered subregions. Following their approach, the POS magnetic field strength in the most general form is given by
\begin{equation}\label{DMAEq}
    B_{\mathrm{POS}} \approx f \sqrt{4 \pi \langle\rho\rangle} \frac{D_{2D}^{1/2}\{C\}(\ell)}{D_{2D}^{1/2}\{\theta\}(\ell)},
\end{equation}
where $f$ is a correction factor and $\langle\rho\rangle$ is the mean density of the gas. The expressions $D_{2D}\{C\}(\ell)$ and $D_{2D}\{\theta\}(\ell)$ describe the second-order structure functions of the mean velocity and the polarization angle, respectively, measured at distance $\ell$. For some quantity $Q(\mathbf{X})$ at POS coordinate $\mathbf{X}$, the second-order structure function $D_{\mathrm{2D}}\{Q\}(\mathbf{\ell})$ is given by the expression
\begin{equation}
    D_{\mathrm{2D}}\{Q\}(\mathbf{\ell}) = \langle[Q(\mathbf{X}+\mathbf{\ell})-Q(\mathbf{X})]^2\rangle_\mathbf{X}.
\end{equation}
Similar to the regularly employed DCF method (\citealp{davis_strength_1951,chandrasekhar_problems_1953}) and its more recent modifications (e.g., \citealp{hildebrand_dispersion_2009, Houde_2009, Skalidis_2021}), with DMA, one compares the dispersion of velocities and magnetic field directions. This relation stems from the Alfvénic coupling between perturbations in turbulent velocity $\delta v_{\mathrm{turb}}$ and magnetic field $\delta B$, given by $\delta v_{\mathrm{turb}} = \delta B/\sqrt{4\pi\rho}$. However, \citet{Lazarian_2022} argue that the use of global measures, as is the case with DCF, is only applicable when the sampling size for the cloud is greater than the scale $L_{\mathrm{inj}}$, at which turbulent motions are injected into the system\footnote{For the ISM, $L_\mathrm{inj}$ is believed to be around 100\,pc \citep{Chepurnov_2010}.}. If this is not the case, DCF will overestimate the magnetic field strength.\\
DMA also allows one to account for the anisotropy of MHD turbulence, by encoding the dependence of $B_\mathrm{POS}$ on the Alfvénic Mach number $\mathcal{M}_\mathrm{A}$, the angle between the mean magnetic field and the POS $\gamma$, and the ratio between gas pressure and magnetic pressure $\beta_{\mathrm{gas}/\mathrm{mag}}$ in the correction factor $f = f(\mathcal{M}_\mathrm{A},\gamma,\beta_{\mathrm{gas}/\mathrm{mag}})$. They derive the correction factor for the low $\beta_{\mathrm{gas}/\mathrm{mag}}$-regime through numerical simulations and find
\begin{equation} \label{DMAfeq}
    f(\gamma) \approx 
    \begin{cases}
    0.5 \mathcal{M}_\mathrm{A}^{1/2}, \qquad \qquad\mathrm{if} \quad \gamma = \pi/2\\
    0.83 \mathcal{M}_\mathrm{A}^{1/2}, \qquad \quad \;\: \mathrm{if} \quad \gamma = \pi/4 
    \end{cases}
\end{equation}

\noindent (see their Figs. 15 and 20). Inferring the angle $\gamma$ is not straightforward. Therefore, we will use these values as our upper and lower bounds for the estimation of $B_{\mathrm{POS}}$.
\subsubsection{Velocity structure}
DMA makes use of velocity centroids, available from spectroscopic observations of Doppler-shifted molecular line transitions, as a proxy for the spatial distribution of the mean velocity. Given the Intensity $I(\mathbf{X},v_i)$ of the molecular line in the $i$-th velocity channel $v_i$ and at POS coordinate $\mathbf{X}$, one can find the velocity centroid $C(\mathbf{X})$ by computing
\begin{equation}
    C(\mathbf{X}) = \sum_i v_i I(\mathbf{X},v_i)/ \sum_i I(\mathbf{X},v_i)
\end{equation}
over a chosen molecular line. One caveat of this approach is that the line profile is not solely influenced by velocity fluctuations along the LOS, but also by fluctuations in density. We aim to mitigate this issue by decomposing the channel intensity into a contribution that is correlated with density (which we disregard for our analysis) and one that is not (for details, we refer to \citealp{Yuen_2021}).\\
We compute density-corrected velocity centroids of the $\mathrm{N_2H^+}(101-012)$ hyperfine transition from the data described in Sect.~\ref{NROKap} over a range of 3~km~s$^{-1}$. $\mathrm{N_2H^+}$ has proven to be a useful tracer of gas kinematics inside dense cloud cores, as it is less affected by condensation on dust surfaces compared to CO and remains optically thin even at high volume densities (\citealp{Tafalla_2002,Lippok_2013}). The resulting centroid map is then regridded to a resolution of $12^{\prime\prime}$, according to the procedure described in Sect.~\ref{DataPrepKap}. Given the larger beam size of the NRO data (see Table~\ref{InstrumentsTab}), for this section, we reapply Eq.~\ref{smoothingEq} to the polarization data, as we need to make sure the resolution of all datasets remains comparable\footnote{The selection of polarization vectors, as shown in Fig.~\ref{FluxMapFig}, remains the same as in the previous sections.}. \\
The study of velocity centroids also allows us to find an independent measure for the Alfvénic Mach number $\mathcal{M}_\mathrm{A}$ through the application of the gradient technique (GT, \citealp{Gonzales_2017,Lazarian_2018,Lazarian_2024}), where gradients of velocity centroids are used to infer properties of the underlying magnetic field. \citet{Lazarian_2024} derive a connection between the gradient covariance tensor along the directions $i$ and $j$, i.e., $\sigma_{ij} \equiv \langle\nabla_{X^i}C(\mathbf{X})\nabla_{X^j}C(\mathbf{X})\rangle$ and $\mathcal{M}_\mathrm{A}$, given by the expression
\begin{equation}\label{M_AEq}
    \frac{(\sigma_{xx}-\sigma_{yy})^2+4\sigma_{xy}}{(\sigma_{xx}+\sigma_{yy})^2}\approx \exp(-2(\mathcal{M}_\mathrm{A} / \sin \gamma)^2).
\end{equation}

\subsubsection{Application of DMA}
\begin{figure}
  \centering
  \includegraphics[width=\columnwidth]{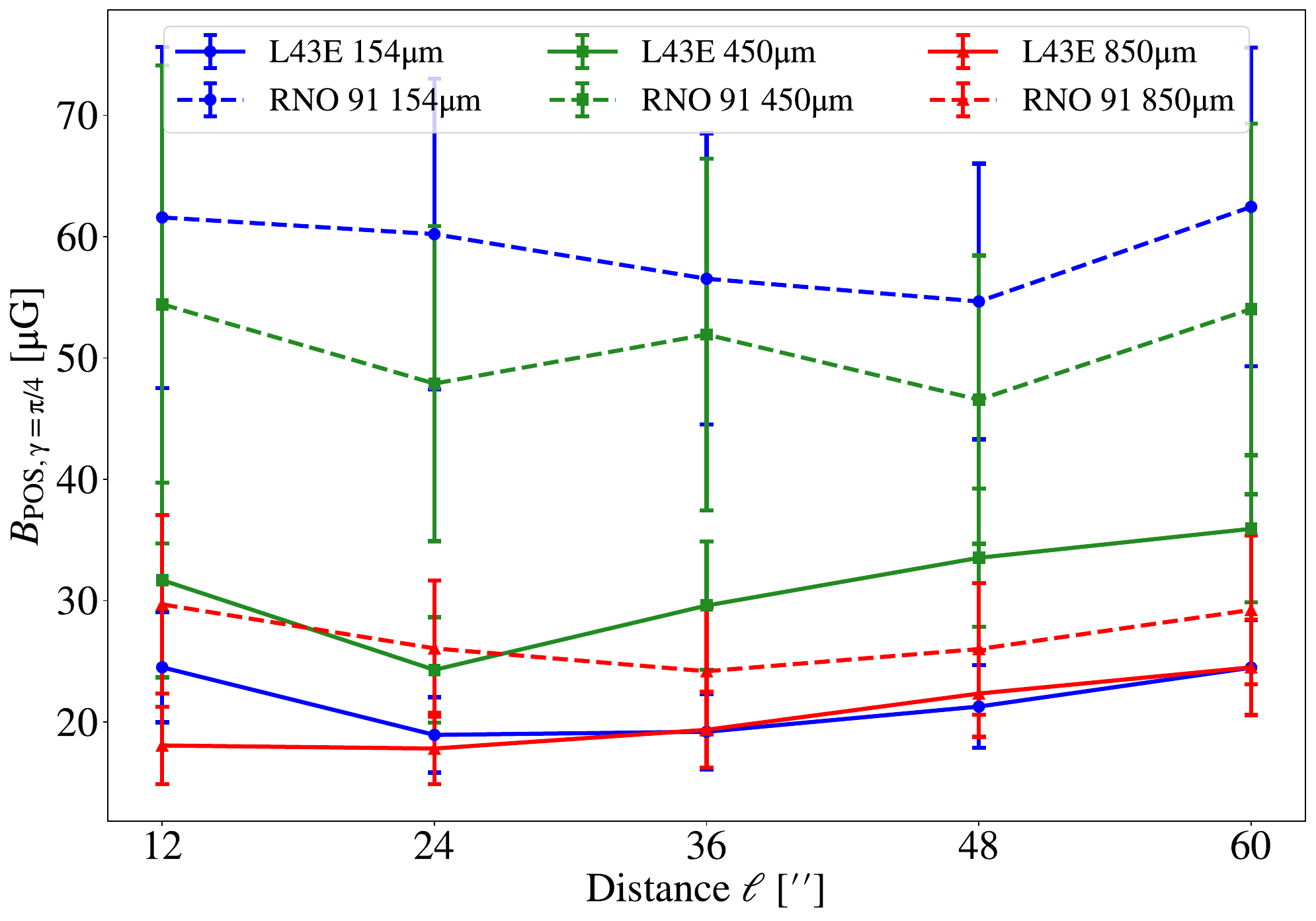}
  \caption{POS magnetic field strength $B_\mathrm{POS, \gamma = \pi/4}$ as a function of the separation distance $\ell$ for the L43E (solid lines) and RNO 91 (dashed lines) regions. Polarization measurements are taken at 154$\,\mu\mathrm{m}$ (blue), 450$\,\mu\mathrm{m}$ (green) and 850$\,\mu\mathrm{m}$ (red).}
  \label{DMAFig}
\end{figure}

We estimate the mean density $\langle\rho\rangle$ in the two subregions assuming that all dust and gas along the LOS is located inside an ellipsoid with semimajor and semiminor POS half-axes $a_\mathrm{HA}$ and $b_\mathrm{HA}$ and the depth of the ellipsoid being taken as the geometric mean $c_\mathrm{HA} = \sqrt{a_\mathrm{HA}b_\mathrm{HA}}$ of these half-axes \citep{karoly_jcmt_2023}. The mean volume density for each subregion is then given by
\begin{equation}
    \langle\rho\rangle = \mu m_\mathrm{H}n_\mathrm{H_2}= \frac{3\mu m_\mathrm{H}}{4}\frac{N_\mathrm{H_2}}{ c_\mathrm{HA}}[\mathrm{g\,cm^{-3}}].
\end{equation}
By combining Eqs.~\ref{DMAfeq} and \ref{M_AEq}, one obtains the value of $f$ for the two considered inclinations of the magnetic field. To obtain an estimate of $B_\mathrm{POS}$, we calculate the structure functions $D_{2D}^{1/2}\{C\}(\ell)$ and $D_{2D}^{1/2}\{\theta\}(\ell)$ at a separation distance $\ell = 24^{\prime\prime}$. This distance ensures the independence of the data pairs, as it is larger than the largest beam size, and likely also fulfills the second criterion for the application of DMA, which is that $\ell$ is required to be smaller than the LOS extent of the cloud \citep{Lazarian_2022}. Fig.~\ref{DMAFig} shows the values of $B_\mathrm{POS,\gamma = \pi / 4}$, calculated from Eq.~\ref{DMAEq}, for each subregion and wavelength as a function of the distance $\ell$ between the data pairs. \\
Given the POS magnetic field strength and the column density (see Sect.~\ref{BbfitsKap}), one can use the mass-to-flux ratio $\lambda_{M/\Phi}$ (\citealp{nakano_gravitational_1978,crutcher_scuba_2004})
\begin{equation}\label{MasstofluxEq}
    \lambda_{M/\Phi} = \frac{(M/\Phi)_{\mathrm{obs}}}{(M/\Phi)_{\mathrm{crit}}} =  7.6 \times 10^{-21} \frac{N_{\mathrm{H_2}}}{B_{\mathrm{POS}}}\left\lbrack\frac{\mathrm{cm^{-2}}}{\mu\mathrm{G}}\right\rbrack
\end{equation}

\noindent to obtain a measure for the required magnetic flux $\Phi$ that would prevent the gravitational collapse of a structure with mass $M$. If $\lambda_{M/\Phi} < 1$, the magnetic field is strong enough to prevent the structure from collapsing (magnetically subcritical regime), while $\lambda_{M/\Phi} > 1$ indicates that the magnetic field is unable to prevent the collapse by itself (magnetically super-critical regime). The determination of $\lambda_{M/\Phi}$ through this method suffers from geometrical biases, depending on the LOS and the orientation of the cloud, as explained in~\citet{crutcher_scuba_2004}. A statistical correction for this bias yields $\lambda_{M/\Phi,\mathrm{cor}}=\lambda_{M/\Phi}/3$. A summary of all results obtained in this section is given in Table~\ref{DMATab}. A discussion of the results can be found in Sect.~\ref{StabilityKap}.\\

\begin{table*}
  \renewcommand{\arraystretch}{1.2}
  \setlength{\tabcolsep}{6pt}
  \centering
  \caption{Results of the DMA analysis.}
  \label{DMATab}  
  \begin{tabular}{c *{3}{r@{\,$\pm$\,}l}@{\hspace{1cm}}*{3}{r@{\,$\pm$\,}l}}
    \hline\hline
    \noalign{\vskip 1pt}
    \raisebox{-5pt}[0pt][0pt]{Quantity}
      & \multicolumn{6}{c}{L43E region} 
      & \multicolumn{6}{c}{RNO 91 region} \\
    \cline{2-7}\cline{8-13}
    \noalign{\vskip 1pt}
      & \multicolumn{2}{c}{154\,$\mu$m}
      & \multicolumn{2}{c}{450\,$\mu$m}
      & \multicolumn{2}{c}{850\,$\mu$m}
      & \multicolumn{2}{c}{154\,$\mu$m}
      & \multicolumn{2}{c}{450\,$\mu$m}
      & \multicolumn{2}{c}{850\,$\mu$m} \\
    \noalign{\vskip 1pt}
    \hline
        $a_\mathrm{HA}[^{\prime\prime}]$
      & 90\,\,\, & \,\,\,1 & 90\,\,\, & \,\,\,1 & 90\,\,\, & \,\,\,1
      & 70\,\,\, & \,\,\,1 & 70\,\,\, & \,\,\,1 & 70\,\,\, & \,\,\,1 \\
          $b_\mathrm{HA}[^{\prime\prime}]$
      & 59\,\,\, & \,\,\,1 & 59\,\,\, & \,\,\,1 & 59\,\,\, & \,\,\,1
      & 50\,\,\, & \,\,\,1 & 50\,\,\, & \,\,\,1 & 50\,\,\, & \,\,\,1 \\
    $n_{\mathrm{H_2}}$ [$10^4\,\mathrm{cm^{-3}}$]
      & 7.6\,\,\, & \,\,2.0 & 7.6\,\,\, & \,\,2.0 & 7.6\,\, & \,\,2.0
      & 5.8\,\,\, & \,\,1.9 & 5.8\,\, & \,\,1.9 & 5.8\,\, & \,\,1.9 \\
    $\mathcal{M}_\mathrm{A}(\gamma=\pi/2)$
      & 0.96 & \, 0.06&  0.96& \, 0.06&  0.96& \,\,0.06
      & 1.03 & \, 0.09&  1.03& \, 0.09&  1.03& \,0.09 \\

    $\mathcal{M}_\mathrm{A}(\gamma=\pi/4)$
      &  0.69& \, 0.06&  0.69& \,\,0.06 &  0.69& \,\,0.06
      &  0.73& \, 0.09&  0.73& \, 0.09&  0.73& \,0.09 \\

    $f(\gamma=\pi/2)$
      & 0.49 & \,\,0.03 &  0.49& \,\,0.03 &  0.49& \,\,0.03
      &  0.51& \,\,0.04 &  0.51& \,\,0.04 &  0.51& \,0.04 \\

    $f(\gamma=\pi/4)$
      &  0.68& \,\,0.06 &  0.68& \,\,0.06 &  0.68& \,\,0.06
      &  0.71& \, 0.09& 0.71 & \,\,0.09 &  0.71& \,0.09 \\

    $B_{\mathrm{POS,\gamma=\pi/2}}\,[\mu\mathrm{G}]$
      & 13.6\,\,\, & \,\,2.0 &  17.4\,\,\,& \,\,2.9 & 12.8\,\,\, & \,\,1.9
      & 43.1\,\,\, & \,\,\,8.3 & 34.3\,\,\, & \,\,\,8.8 & 18.7\,\,\, & \,3.7 \\

    $B_{\mathrm{POS,\gamma=\pi/4}}\,[\mu\mathrm{G}]$
      & 18.9\,\,\, & \,\,3.1 & 24.3\,\,\, & \,\,4.3 & 17.8\,\,\, &  \,\,2.9
      &60.2\,\,\, & 12.7 & 47.9\,\,\, & 12.9 & 26.1\,\,\, &  \,5.6 \\
      
    $\lambda_{M/\Phi,\mathrm{cor},\gamma=\pi/2}$
      & 2.58 & \,\,0.75 & 2.01 & \,\,0.60 & 2.74 & \,\,0.79
      & \multicolumn{2}{c}{--}
      & \multicolumn{2}{c}{--}
      & \multicolumn{2}{c}{--} \\

    $\lambda_{M/\Phi,\mathrm{cor},\gamma=\pi/4}$
      & 1.85 & \,\,0.55 & 1.44 & \,\,0.44 & 1.96 & \,\,0.58
      & \multicolumn{2}{c}{--}
      & \multicolumn{2}{c}{--}
      & \multicolumn{2}{c}{--} \\

    \hline
  \end{tabular}
\end{table*}

\subsubsection{Comparison with DCF}
The DCF technique has been the primary method for inferring magnetic field strengths in the ISM for many years. To test the prediction made by DMA, namely that DCF will overestimate the magnetic field strength for $\ell \ll L_{\mathrm{inj}}$, we also applied the modified DCF approach of \citet{hildebrand_dispersion_2009} to our data. We find that for both subregions, the DCF estimates are higher by a factor of $\sim 1.5 - 2.5$, in agreement with the prediction of \citet{Lazarian_2022}. 

\section{Modeling the polarization spectrum of a molecular cloud core with POLARIS}\label{PolarisKap}
\subsection{The model}\label{ModelKap}
In this Section, we employ the 3D Monte Carlo radiative transfer code POLARIS \citep{reissl_radiative_2016} to analyze how variations in dust alignment properties and temperatures along the LOS affect the slope of the polarization spectrum under the radiative torque paradigm. For this purpose, we construct a simple model of a molecular cloud core that is irradiated by the interstellar radiation field (ISRF), as well as a nearby star. We assume a Bonnor-Ebert density distribution of the form

\begin{equation}
\rho(r) =
\begin{cases}
\rho_0, \qquad\quad\;\;\,\mathrm{if} & r \leq r_0,\\
\rho_0 r^{-2}/r_0, \quad\,\mathrm{if} & r_0 < r \leq r_{\mathrm{out}},\\
0, \qquad\qquad\,\mathrm{if} & r_{\mathrm{out}} < r,
\end{cases}
\end{equation}
which has already been used in other studies to model molecular cloud cores (e.g., \citealp{brauer_origins_2016,zielinski_constraining_2021,giang_physical_2023}). Here, $r_0$ is the truncation radius, at which the density distribution begins to decrease with $r^{-2}$, until the outer radius $r_{\mathrm{out}}$ is reached. The model parameters (see Table~\ref{ModelParamTab}) are chosen to resemble the environment of L43. Based on the emission profiles in Fig.~\ref{FluxMapFig} and the distance estimate of \citet{de_geus_physical_1989}, we assume a truncation radius of $3000\,\mathrm{au}$, an outer radius of $3\times10^{4}\,\mathrm{au}$, and a distance between the YSO and the density peak of L43E of $10^{4}\,\mathrm{au}$. The total mass is taken to be $3\,\mathrm{M_\odot}$, matching the previously derived column density in L43 (see Fig.~\ref{BBFitFig}). We consider two different setups. In Setup one, we only consider the ISRF, where photon packages start propagating from the outside of the model space. The total energy density of the ISRF is taken to be $u_{\mathrm{ISRF}}=8.64\times10^{-13}\,\mathrm{erg\,cm^{-3}}$, which is equal to the total energy density of the solar neighborhood \citep{mathis_interstellar_1983}. Setup two considers the ISRF, as well as a stellar source, located inside the model space at a displacement radius of $10^{4}\,\mathrm{au}$. We apply a constant, homogeneous magnetic field $B=100\,\mu\mathrm{G}$, aligned perpendicular to the LOS and rotated by $150^\circ$ with respect to the axis defined by the center of the model space and the location of the star in Setup two, to roughly match the large-scale field of L43 seen by Planck~\citep{ade_planck_2015}. The energy of the photon packages propagating from the star is given by its SED, which is modeled to resemble RNO 91. We assume a surface temperature of $4000\,\mathrm{K}$, motivated by the average effective temperatures of young stars in the region (\citealp{doppmann_physical_2005,mcclure_evolutionary_2010}) with a bolometric luminosity of $L_{\star}=2.5 \mathrm{L_\odot}$ \citep{chen_spitzer_2009}. We consider oblate dust grains with effective radii $a$ ranging from $5\,\mathrm{nm}$ to $500\,\mathrm{nm}$ and an axis ratio of $1:2$, distributed in the form of a power law ($n(a) \propto a^{\alpha}$), with $\alpha = -3.5$. The dust consists of a mixture of $62.5\%$ silicate and $37.5\%$ carbonaceous grains, whose optical properties are calculated with DDSCAT \citep{Draine:94} using the optical properties given in \citet{draine_scattering_2003}. As already pointed out by \citet{seifried_origin_2023}, differences in temperatures between aligned and unaligned dust species are a key contributor to the observed shape of the polarization spectrum. For a given density distribution consisting of a composition of dust species, POLARIS calculates a singular dust temperature for each cell by default. Thus, we consider two distinct but spatially overlapping dust density distributions for each dust species. Guidance on the implementation of this feature is given in Appendix A of \citet{seifried_origin_2023}. Size-dependent dust temperatures are not considered. Instead, we estimate their impact in Appendix~\ref{AppB}. We assume a constant gas-to-dust mass ratio of $100:1$.\\

\begin{table}
\caption{Model parameters.}
\label{ModelParamTab}
\centering
\begin{tabular}{l c c}
\hline\hline
Parameter                                  & Symbol                 & Value             \\
\hline
\textbf{\underline{Structure:}}            &                        &                   \\[1mm]
Truncation radius                          & $r_{0}$                & $3000\,\mathrm{au}$   \\
Outer radius                               & $r_{\mathrm{out}}$     & $3\times10^4\,\mathrm{au}$  \\
Gas mass                                   & $M_{\mathrm{gas}}$     & $3\,\mathrm{M_\odot}$  \\
Magnetic field strength                    & $B$     & $100\,\mu\mathrm{G}$  \\
Distance                                   & $d$                    & $125\,\mathrm{pc}$   \\
\textbf{\underline{Heating sources:}}      &                       &                  \\[1mm]
\textbf{Star:}                             &                       &                  \\
Luminosity                                 & $L_\star$            &   $2.5\,\mathrm{L_\odot}$ \\
Surface temperature                        & $T_\star$            &   $4000\,\mathrm{K}$ \\
Distance from core center                  & $r_{\mathrm{disp}}$  &   $10^4\,\mathrm{au}$     \\
\textbf{ISRF:}                             &                      &                    \\
Total energy density                       & $u_{\mathrm{ISRF}}$    &  $8.64\times10^{-13}\,\mathrm{erg\,cm^{-3}}$                 \\
\textbf{\underline{Dust properties:}}      &                        &                  \\[1mm]
Minimum grain size                         & $a_{\min}$             &    $5\,\mathrm{nm}$     \\
Maximum grain size                         & $a_{\max}$             &    $500\,\mathrm{nm}$    \\
Exponent of the\\grain size distribution        & $\alpha$                    &   $-3.5$    \\
Grain axial ratio                          & $s$                    & $1:2$             \\
Silicate–Carbon ratio                      & $f_{\mathrm{Sil/Carb}}$     &   $1.67:1$     \\
Gas‐to‐dust mass ratio                     & $f_{\mathrm{g/d}}$           &  $100:1$        \\
\textbf{\underline{Alignment properties:}} &                      &                  \\[1mm]
Ratio of grains at a\\ high-J attractor point& $f_{\mathrm{high-J}}$ & 0.25        \\
\hline
\end{tabular}
\end{table}

\subsection{Alignment mechanisms}\label{AligMechKap}
In the following, we assume that the alignment of the grains along their minor axis and the direction of the external magnetic field is given by RAT alignment. In POLARIS, the torque $\Gamma_{\mathrm{rad}}$ exerted on a grain is calculated using the expression (\citealp{hoang_grain_2014,reissl_systematic_2020})
\begin{equation}
    \Gamma_{\mathrm{rad}} = \pi a^2 \int \left(\frac{\lambda}{2\pi}\right) \gamma_{\lambda}\cos(\psi)Q_\Gamma(a,\lambda)u_\lambda\mathrm{d}\lambda.
\end{equation}

\noindent Here, $\gamma_\lambda$ is a factor quantifying the degree of anisotropy of the radiation field. A value of $\gamma_\lambda=1$ indicates a unidirectional radiation field, while $\gamma_\lambda = 0$ occurs if the radiation field is isotropic. The quantity $\psi$ refers to the angle between the direction of the incident radiation and that of the magnetic field, $Q_\Gamma(a,\lambda)$ is a factor determining the RAT efficiency, and $u_\lambda$ is the spectral energy density of the radiation field. Dust grains are assumed to be aligned if the condition
\begin{equation}
    \frac{J_{\mathrm{rad}}}{J_\mathrm{th}} = \frac{\tau_\mathrm{drag}\Gamma_\mathrm{rad}}{J_\mathrm{th}}\geq 3
\end{equation}
is met, where $J_\mathrm{rad}$ and $J_\mathrm{th}$ refer to the angular momentum caused by radiative torques and thermal excitations, respectively. The quantity $\tau_\mathrm{drag}$ refers to the damping time, after which the grain's direction of rotation is randomized through gas collisions.\\
We also consider imperfect internal alignment. Here, internal alignment refers to the ability of the grain to align its angular momentum parallel to the short axis of the grain. This condition depends on the angular momentum a grain receives under the influence of RATs. \citet{lazarian_radiative_2007} found that the angular momentum of a grain settles at either a high-J or low-J attractor point. Grains at high-J are less prone to processes that randomize their orientation and therefore contribute more to the polarized emission than grains sitting at a low-J attractor point. The ratio $f_{\mathrm{high-J}}$ of grains that settle at the high-J attractor point depends on the size and magnetic properties of the grain (\citealp{hoang_unified_2016,giang_physical_2023}), but is still poorly constrained. Therefore, we assume a conservative estimate of $f_{\mathrm{high-J}}=0.25$.   

\subsection{Fitting the spatially resolved polarization spectrum}\label{PolarisMethodsKap}
Polarization maps with a pixel resolution of $255\times255$, corresponding to $3.14^{\prime\prime}\mathrm{px^{-1}}$, are calculated for the observing wavelengths of $154\,\mu\mathrm{m}$, $450\,\mu\mathrm{m}$ and $850\,\mu\mathrm{m}$. The maps are convolved using a Gaussian kernel with a FWHM of $14.6^{\prime\prime}$, corresponding to the effective beam size of the set of polarization observations used in Sect.~\ref{ObsResultsKap}, after Eq.~\ref{smoothingEq} is applied.\\
Keeping in line with previous works (e.g., \citealp{gandilo_submillimeter_2016,michail_far-infrared_2021,lee_modeling_2024,cox_sofia_2025}), we investigate changes in the slope of the polarization spectrum across different LOSs by fitting a function of the form
\begin{equation}\label{FitEq}
    p(\lambda)/p_{\lambda_0}=a_l(b_l[\lambda-\lambda_0]+1),
\end{equation}
to each pixel of the model space. Here, similar to Sect.~\ref{PolSpecKap}, the reference wavelength $\lambda_0$ is taken to be $450\,\mu\mathrm{m}$. Given that $a_l\approx1$\footnote{This holds true as long as $p(\lambda)$ scales linearly with $\lambda$ in the vicinity of $\lambda_0$.}, we investigate correlations between the effective slope $b_l$ and cloud properties, specifically the column density and LOS temperature. 

\subsection{Results}\label{PolarisResultsKap} 
\begin{figure}[htbp]
  \centering
  \includegraphics[width=\columnwidth]{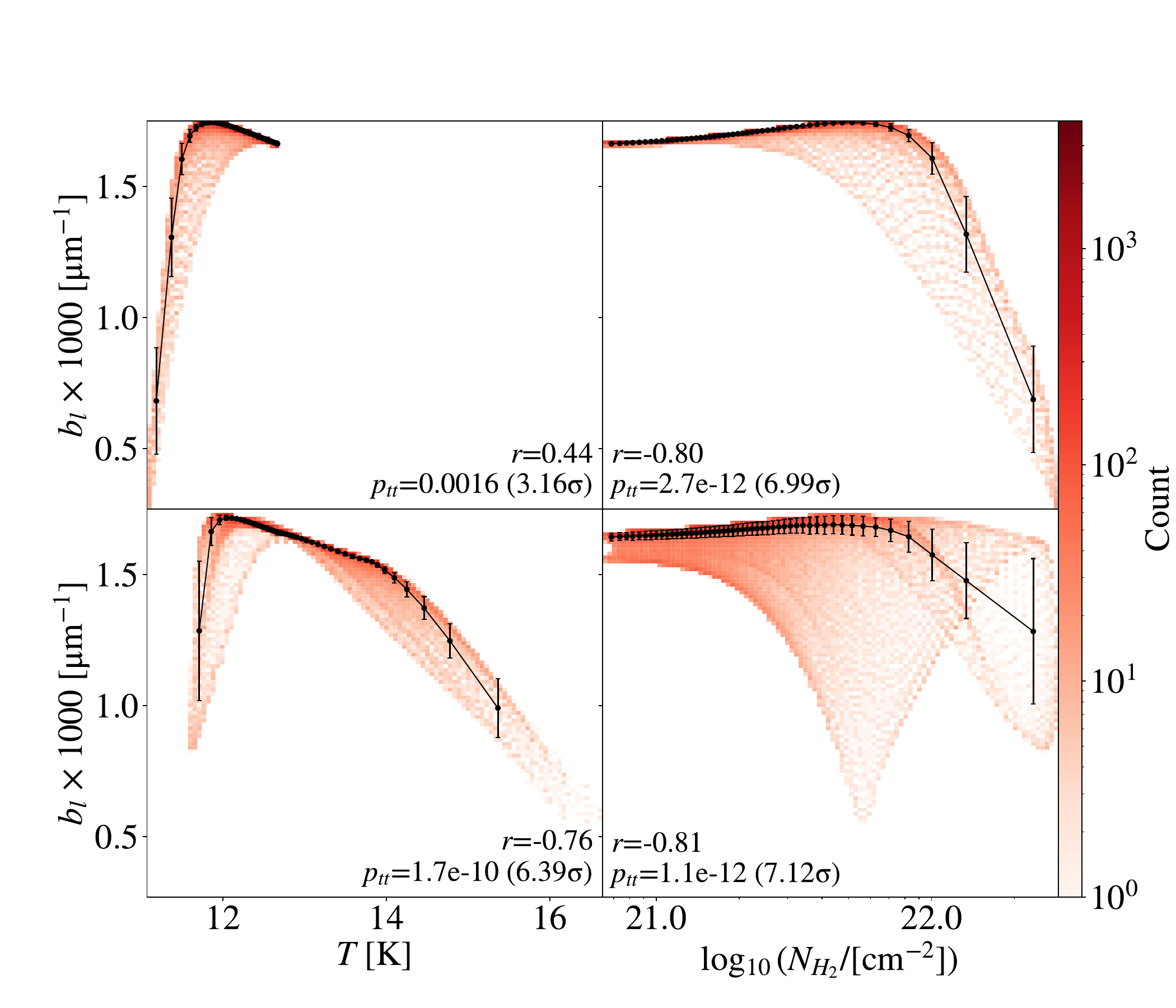}
  \caption{2D histograms of the effective slope $b_l$ of the normalized polarization spectrum, as a function of LOS temperature (left panels) and column density (right panels), respectively, colored by the count of data points in each bin. Top panels show results obtained with the ISRF as the only considered radiation source (Setup one), whereas bottom panels show results where the ISRF as well as an offset stellar source are considered (Setup two, see Sect.~\ref{ModelKap} for details). Pearson $r$, as well as two-tailed $p$ values ($p_{tt}$), are calculated from $50$ segmented median values in each distribution.}
  \label{BlvsTNHFig}
\end{figure}

\begin{figure}[htbp]
  \centering
  \includegraphics[width=0.8\columnwidth]{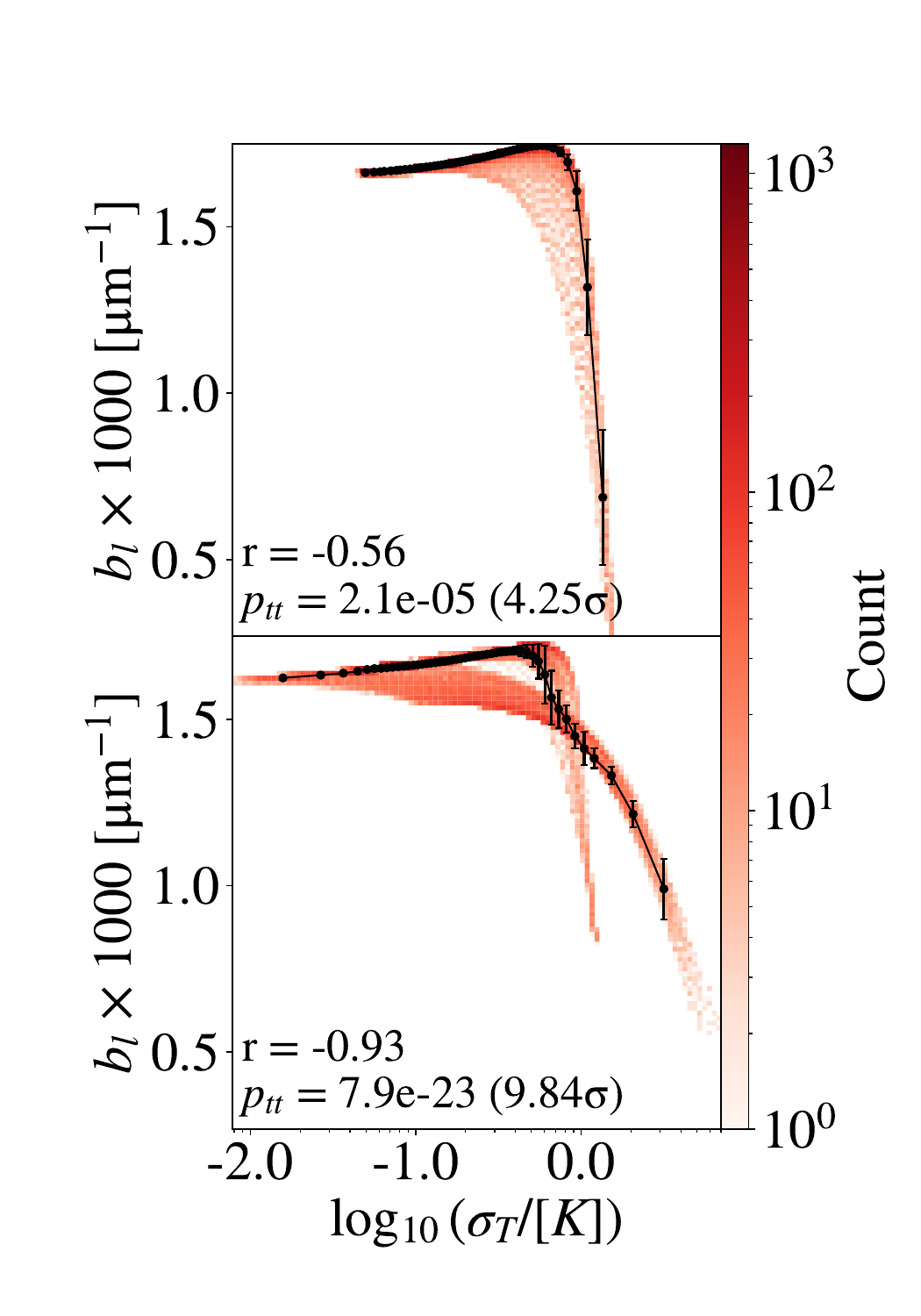}
  \caption{2D histograms of the effective slope $b_l$ of the normalized polarization spectrum, as a function of the standard deviation of the LOS temperature profile $\sigma_T$, colored by the count of data points in each bin. Top panels show results obtained with the ISRF as the only considered radiation source (Setup one), whereas bottom panels show results where the ISRF as well as an offset stellar source are considered (Setup two, see Sect.~\ref{ModelKap} for details). Pearson $r$, as well as two-tailed $p$ values ($p_{tt}$), are calculated from $50$ segmented median values in each distribution.}
  \label{BlvsSigmaTFig}
\end{figure}

In Fig.~\ref{BlvsTNHFig}, the value of the effective slope $b_l$ is shown as a function of LOS temperature and column density. The distribution is segmented into $50$ chunks of equal size. The median and MAD of each chunk are displayed as a point and corresponding error bar, respectively. From these medians, we infer correlations by computing the Pearson $r$ correlation coefficient along with a two-tailed $p$ value, denoted as $p_{tt}$\footnote{The naming of $p_{tt}$ is chosen to avoid confusion with the degree of polarization $p$.}, indicating the probability that the null hypothesis of no correlation is incorrect.\\
For Setup one, where the ISRF acts as the only radiation source (top panels of Fig.~\ref{BlvsTNHFig}), we find LOS temperatures ranging from $13~\mathrm{K}$ in the outer regions of the model space to $11~\mathrm{K}$ in the center regions. This decline is expected, since the radiation responsible for heating the grains is increasingly attenuated as it traverses through the model space. Therefore, due to the symmetry of this setup, column density and LOS temperature are strongly anticorrelated. With Setup one, we find spectra with positive slopes throughout the model space that decline steeply for column densities $\gtrsim 10^{22}\,\mathrm{cm^{-2}}$ or LOS temperatures of $\lesssim12\,\mathrm{K}$. Similar trends have been observed in observational studies (e.g., \citealp{michail_far-infrared_2021,cox_sofia_2025}). However, more diffuse LOSs in the outer regions of the model space exhibit the opposite of this trend, where $b_l$ first increases with increasing column density (or decreasing LOS temperature), but this effect is not of any physical nature. Rather, it is the result of fitting a linear function to a normalized, nonlinear distribution. In Appendix~\ref{AppC}, we further investigate this effect by applying the fits to the data before normalization.\\
In Setup two (bottom rows in Fig.~\ref{BlvsTNHFig}), the influence of the stellar radiation results in higher LOS temperatures. In contrast to Setup one, after an initial increase, $b_l$ decreases with increasing LOS temperatures. Data points on the far side of the stellar source are less affected by the stellar source and thus show a similar dependency to those in Setup one, leading to the initial positive correlation. The data points closest to the stellar source exhibit spectra with the shallowest slopes. To explain these differences in correlation between external and embedded radiation sources, we employ another measure, that is, the standard deviation of the LOS temperature profile $\sigma_T$ \citep{lee_modeling_2024}. This measure aims to quantify the degree of heterogeneity along a given LOS, which is thought to be the cause of the observed negative or V-shaped spectra (\citealp{santos_far-infrared_2019,michail_far-infrared_2021,seifried_origin_2023}). Fig.~\ref{BlvsSigmaTFig} shows the correlation between $b_l$ and $\sigma_T$. In contrast to LOS temperature and column density, $b_l$ decreases with $\sigma_T$ independently of the radiation source considered, indicating that heterogeneity along the LOS governs the slope of the polarization spectrum.


\section{Discussion}\label{DiscussionKap}
\subsection{Polarization efficiency}
The RAT mechanism provides a natural explanation for the observed trend between the degree of polarization and the normalized intensity (Sect.~\ref{PolHoleKap}), in which dust that is shielded from radiation exhibits a lower polarization efficiency compared to dust in a radiation-rich environment, such as the outer layers of molecular clouds. The fact that the polarization efficiency shows a similar decrease in the RNO 91 region, which contains an embedded radiation source, challenges this interpretation. However, when considering only the innermost region around the YSO, we observe a decrease in the value of the fit parameter $a_2$, indicative of a slight but significant enhancement of the polarization efficiency in this region compared to the rest of the cloud.\\
Although some alternative explanations for the occurrence of depolarization toward density peaks, such as the influence of dichroic extinction \citep{brauer_origins_2016} can be ruled out due to the low optical depths at all observing wavelengths (see Sect.~\ref{BbfitsKap}), other mechanisms, such as beam-averaging effects due to changes in the magnetic field smaller than the beam size or changes in grain composition cannot. 

\subsection{Gravitational stability}\label{StabilityKap}
In Sects.~\ref{BbfitsKap} and~\ref{BfieldKap}, we have determined column densities and magnetic field strengths to constrain the ability of the magnetic field to inhibit the gravitational collapse in the starless core L43E. We find peak column densities of $\sim 5\times10^{22}\,\mathrm{cm^{-2}}$, which are approximately six times lower than what \citet{karoly_jcmt_2023} find, who used a similar dataset and method, but agrees well with the findings of \citet{chen_spitzer_2009}. This discrepancy is likely caused by the fact that \citet{karoly_jcmt_2023} fit temperature and column density in two steps, whereas in our method, we fit them simultaneously. Furthermore, the poorly constrained dust properties used in the fit cause additional uncertainties of up to $50\%$ (\citealp{roy_reconstructing_2014,zielinski_limitations_2024}).\\
The differences in polarization angles observed at various positions, as well as their variations between observing wavelengths, indicate a complex magnetic field structure in the region. The alignment of the inferred magnetic field lines with the cavity walls of the CO outflow seen at $450\,\mu\mathrm{m}$ and $850\,\mu\mathrm{m}$ indicates that the magnetic field in the colder and denser regions of L43E has been influenced by the outflow, further complicating the determination of magnetic field strengths. The absence of this influence in the $154\,\mu\mathrm{m}$ data suggests that we are measuring different parts of the cloud along the LOS, caused by variations in local temperatures. 
The mass-to-flux ratio $\lambda_{M/\Phi,\mathrm{cor}}$ (see Eq.~\ref{MasstofluxEq}), corrected for geometrical biases \citep{crutcher_scuba_2004}, is greater than unity in L43E (see Table~\ref{DMATab}), suggesting that the cloud is likely gravitationally unstable. L43E is slightly sub-Alfvénic, indicating that the dynamics are dominated to a greater extent by the magnetic field, rather than by turbulence. For the RNO 91 region, we do not provide mass-to-flux ratios because the region already contains a YSO. The Alfvénic Mach number in this region slightly exceeds unity when assuming that the mean magnetic field direction lies perpendicular to the LOS. The low dispersion seen especially in the $154\,\mu\mathrm{m}$ data suggests that the dynamics are dominated by the magnetic field, though this might not hold true for all regions within the cloud, as suggested by the high dispersion measured at $850\,\mu\mathrm{m}$.

\subsection{Influences on the shape of the polarization spectrum}\label{InfluencesKap}
The simple model we employed to study the polarization spectrum of a molecular cloud core under a homogeneous magnetic field (see Sect.~\ref{PolarisKap}) shows that an increase in heterogeneity along a given LOS significantly decreases the observed slope of the polarization spectrum -- a phenomenon known as the heterogeneous cloud effect (HCE, \citealp{michail_far-infrared_2021}).\\
In the diffuse regions of the model, the spectrum steadily increases, resulting from differences in temperatures between silicate and carbonaceous grains, as predicted by~\citet{seifried_origin_2023}. Under the assumption that the dust properties, radiation field strength, and column densities that we used in our model are indeed similar to those found in L43, additional influences, such as differences in the inclination of the magnetic field lines along the LOS, are needed to explain the observed negative slopes in Fig.~\ref{PolSpecFig}. The existence of such differences in inclination is plausible, considering that the inclination of the CO outflow is thought to be $\sim 60^{\circ}$ with respect to the POS \citep{lee_outflow_2005}. Given an accurate characterization of the dust properties in a molecular cloud, one can infer the underlying 3D magnetic field structure through multiwavelength polarimetry (e.g., \citealp{tram_understanding_2024}).

\section{Conclusions}\label{ConclusionsKap}
In this paper, we have analyzed the polarization signal in the molecular cloud core L43, arising from magnetically aligned dust grains in the $154-850\,\mu\mathrm{m}$ wavelength range and compared our findings with radiative transfer simulations of a simple cloud core model. Our findings are summarized as follows:\\
\begin{enumerate}
    \item SOFIA/HAWC+ measurements at $154~\,\mu\mathrm{m}$, as well as SCUBA-2/POL-2 measurements at $450\,\mu\mathrm{m}$ and $850\,\mu\mathrm{m}$ reveal wavelength-dependent changes in emission structure, polarization angles and degrees in both considered subregions of the molecular cloud core L43, indicating a complex magnetic field morphology.
    \item The CO outflow associated with the YSO RNO 91 seems to have influenced the polarization signal at $450\,\mu\mathrm{m}$ and $850\,\mu\mathrm{m}$, indicating the alignment of the magnetic field with the outflow direction. This influence is not seen in the $154\,\mu\mathrm{m}$ data.
    \item The relationship between the degree of polarization and the normalized intensity suggests that the polarization efficiency is heavily attenuated in the inner parts of both subregions. However, analysis constrained to the 20$^{\prime\prime}$ region around the YSO suggests that alignment is significantly more retained here, in line with predictions that RATs cause more efficient alignment in the presence of an anisotropic radiation source.
    \item We find peak column densities of $N_{\mathrm{H_2,max}} = (4.72\pm1.25)\,10^{22}\,\mathrm{cm^{-2}}$ in the main starless core L43E, which is considerably lower than what \citet{karoly_jcmt_2023} find using a similar method, but agrees well with previous estimates of \citet{chen_spitzer_2009}. Temperatures in L43E range from $10-15\,\mathrm{K}$. The peak temperature is found in the southern part of the RNO 91 region, giving $T_{\mathrm{d,max}}=(23.5\pm3.9)\,\mathrm{K}$.
    \item We apply the differential measure analysis method \citep{Lazarian_2022} to infer magnetic field strengths in both subregions at all three wavelengths. In the L43E region, we find POS field strengths ranging from $13-24\,\mu\mathrm{G}$, and in the RNO 91 region, we find field strengths ranging from $19-60\,\mu\mathrm{G}$. These estimates are subject to several uncertainties and should be interpreted with care. We find mass-to-flux ratios $\lambda_{M/\Phi, \mathrm{cor}}$ exceeding unity in L43E, suggesting a gravitationally unstable core.
    \item We find polarization spectra with negative slopes throughout the cloud. A comparison with radiative transfer simulations suggests that HCE does have a significant influence on the spectral slope, but additional influences, such as varying inclinations of the magnetic field lines along the LOS, are likely needed to fully explain observations.

\end{enumerate}

\begin{acknowledgements}
The authors thank the anonymous referee for helpful comments and suggestions. We also thank Dr. Janik Karoly for helpful insights on the data reduction with SMURF and Dr. Moritz Lietzow-Sinjen, as well as Dr. Stefan Reissl, for their helpful comments on working with POLARIS. M.S. and S.W. thank the DFG for financial support under grant WO 857/25-1.
This research is based (in part) on observations made with the NASA/DLR Stratospheric Observatory for Infrared Astronomy (SOFIA). SOFIA is jointly operated by the Universities Space Research Association, Inc. (USRA), under NASA contract NNA17BF53C, and the Deutsches SOFIA Institut (DSI) under DLR contract 50 OK 2002 to the University of Stuttgart.
These observations were obtained (in part) by the James Clerk Maxwell Telescope, operated by the East Asian Observatory on behalf of The National Astronomical Observatory of Japan; Academia Sinica Institute of Astronomy and Astrophysics; the Korea Astronomy and Space Science Institute; the National Astronomical Research Institute of Thailand; Center for Astronomical Mega-Science (as well as the National Key R\&D Program of China with No. 2017YFA0402700). Additional funding support is provided by the Science and Technology Facilities Council of the United Kingdom and participating universities and organizations in the United Kingdom and Canada.
Additional funds for the construction of SCUBA-2 were provided by the Canada Foundation for Innovation. 
The Nobeyama 45-m radio telescope is operated by Nobeyama Radio Observatory, a branch of the National Astronomical Observatory of Japan. 
The authors wish to recognize and acknowledge the very significant cultural role and reverence that the summit of Mauna Kea has always had within the indigenous Hawaiian community.  We are most fortunate to have the opportunity to conduct observations from this mountain.\\
\textit{Facilities:} SOFIA (HAWC+), JCMT (SCUBA-2, POL-2), Herschel (PACS, SPIRE)\\
\textit{Software:} \texttt{numpy}\,\citep{van_der_walt_numpy_2011}, \texttt{matplotlib}\,\citep{hunter_matplotlib_2007}, \texttt{scipy}\,\citep{virtanen_scipy_2020}, \texttt{astropy}\,\citep{astropy_collaboration_astropy_2013,astropy_collaboration_astropy_2018}, \texttt{emcee}\,\citep{foreman-mackey_emcee_2013}, \texttt{Starlink}\,\citep{currie_starlink_2014}
\end{acknowledgements}

\bibliographystyle{aa}

\bibliography{bib_L43.bib} 

\appendix

\begin{strip}
\vspace{-1.5cm}
\section{Histograms of polarization degree and angle}
\centering
  \includegraphics[width=\textwidth]{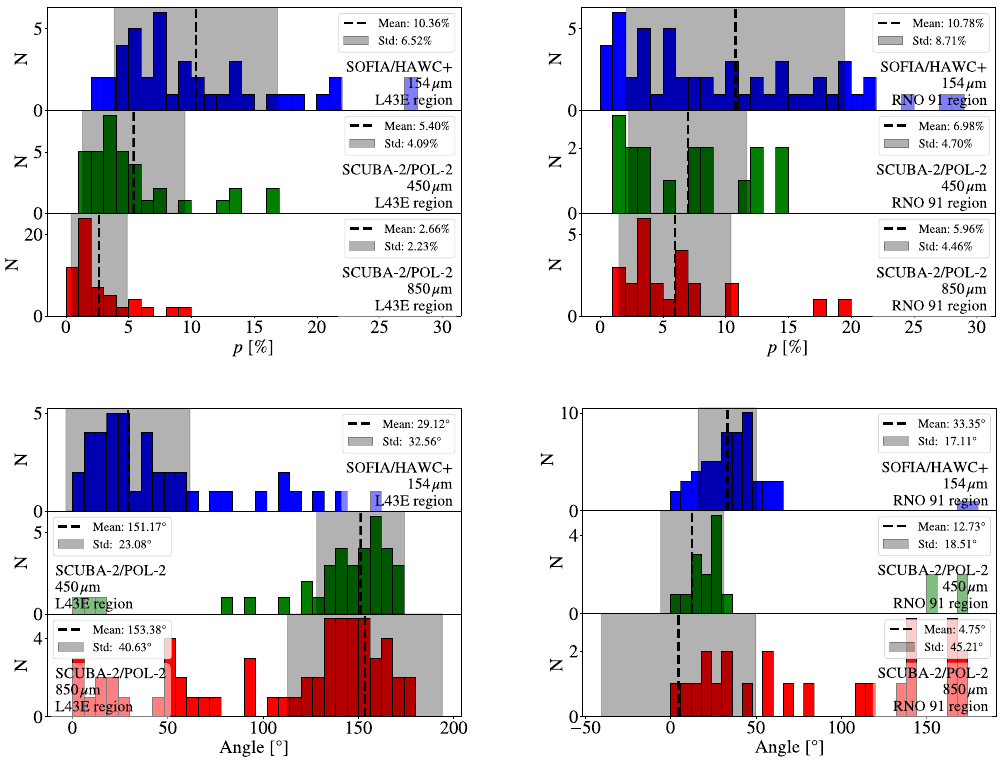}
  \captionof{figure}{Histograms of the degree of polarization (top panels) and the by $90^\circ$ rotated polarization angles (bottom panels) for the three observing wavelengths at $154\,\mu\mathrm{m}$ (blue), $450\,\mu\mathrm{m}$ (green) and $850\,\mu\mathrm{m}$ (red). The dashed black line indicates the mean, and the gray shaded area the standard deviation of the distribution. Left panels: L43E region. Right panels: RNO 91 region.}
  \label{PThetaHistFig}
\end{strip}

\section{The impact of grain-size-dependent dust temperatures on the polarization spectrum}\label{AppB}
\begin{figure*}
  \centering
  \includegraphics[width=\textwidth]{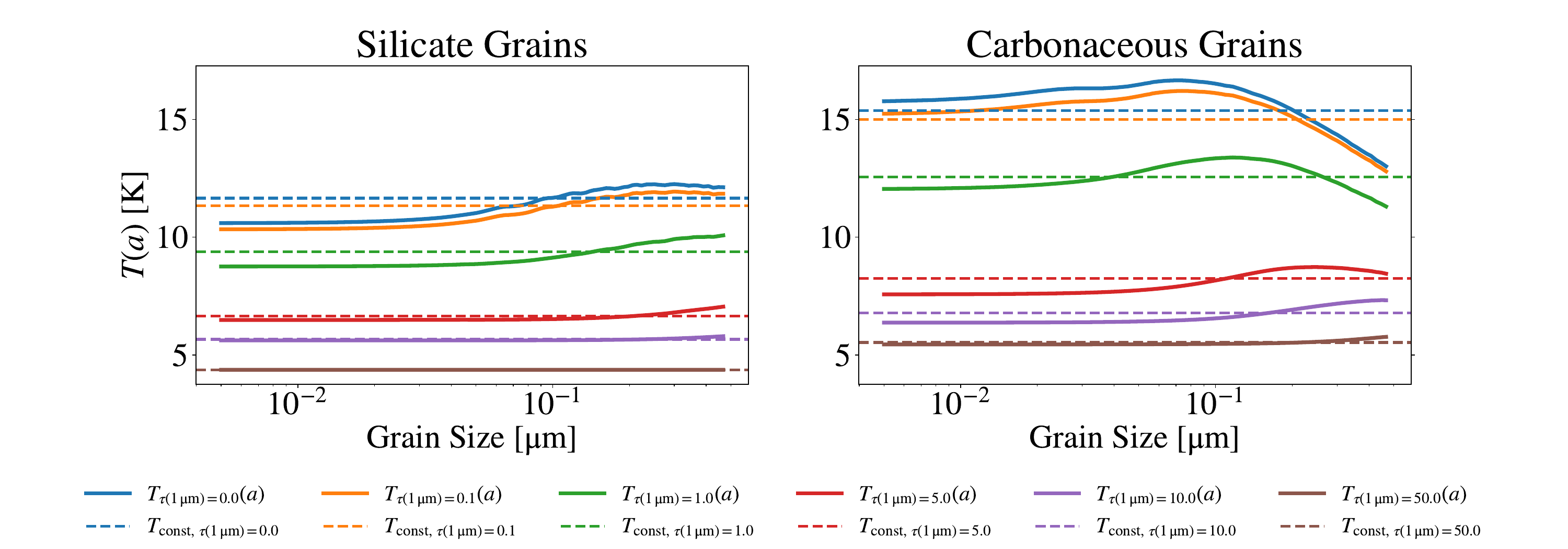}
  \caption{Grain-size-dependent temperatures for differing levels of attenuation (solid lines) along with their mass-weighed averages (dashed lines), calculated with Eq.~\ref{GrainTempEq}. Left: Silicate grains. Right: Carbonaceous grains.}
  \label{AttenuationTempFig}
\end{figure*}
In this Section, we estimate the impact of grain-size-dependent dust temperatures on the shape of the polarization spectrum. In contrast to Sect.~\ref{PolarisKap}, we do not consider the full radiative transfer of individual photon packages. Instead, we calculate the temperature $T_\mathrm{d}(a)$ of a dust grain when subjected to a radiation field with a known spectral energy density $u_\lambda$, assuming local thermal equilibrium. Equating the heating and cooling rates of a grain yields \citep{draine_optical_1984}
\begin{equation}\label{GrainTempEq}
    \frac{4\pi}{c}\int^{\infty}_0 Q_{\mathrm{abs}}(a,\lambda) B_\lambda(T_\mathrm{d}(a))\mathrm{d}\lambda = \int^{\infty}_0 Q_{\mathrm{abs}}(a,\lambda) u_\lambda \mathrm{d}\lambda,
\end{equation} 
where $Q_{\mathrm{abs}}(a,\lambda)$ describes the absorption efficiency for a given grain size $a$ and wavelength $\lambda$. The left side of this equation characterizes the cooling of the grain by the thermal radiation it emits, whereas the right side describes the heating through the absorption of light from the incident radiation field. Given the precalculated absorption efficiencies used in the modeling in Sect.~\ref{ModelKap}, combined with the spectral energy density of the ISRF in the solar neighborhood \citep{mathis_interstellar_1983}, we can calculate the grain-size-dependent temperatures for both of the dust species we consider in Sect.~\ref{PolarisKap}.\\
As the radiation field interacts with dust in a molecular cloud, back-scattering and absorption events gradually attenuate the radiation field, causing the spectral energy density to redden. The grain-size-dependent temperature distribution of an ensemble of dust grains exposed to a reddened radiation field, therefore, changes its shape as well. In Fig.~\ref{AttenuationTempFig}, we use Eq.~\ref{GrainTempEq} to calculate the grain-size-dependent temperatures for silicate grains (left) and carbonaceous grains (right), as well as their mass-weighted average temperatures, for varying optical depths at a reference wavelength of $1\,\mu\mathrm{m}$, given the optical properties of the dust used in Sect.~\ref{PolarisKap}. It should be noted that this approach, unlike the modeling with POLARIS in Sect.~\ref{PolarisKap}, neglects the influence of heating through thermal reemission of dust grains, as well as changes in the shape of the local spectral energy density due to scattering. However, the dust temperatures shown in Fig.~\ref{AttenuationTempFig} correspond to grain-size-dependent temperatures calculated with POLARIS within $0.1\,\mathrm{K}$.\\
To investigate the impact of grain-size-dependent dust temperatures on the polarization spectrum, we employ the semi-analytical model from \citet{seifried_origin_2023}. In this model, changes in the local dust temperature along a given LOS are approximated by two independent phases, each phase contributing to the polarized-, as well as the total intensity at a given wavelength. The total intensity of the cold phase is given by

\begin{equation}
    I_{\mathrm{cold}} = \sum_i f_i \times \int^{a_{\mathrm{max}}}_{a_\mathrm{min}} B_\lambda(T_{\mathrm{d,}i}(a)) \times\pi a^2\times Q_{\mathrm{abs,}i}(a,\lambda)\times a^{\alpha}\mathrm{d}a,
\end{equation}
where $i$ denotes the dust species with mass fraction $f_i$, and $\pi a^2 \times Q_{\mathrm{abs}}(a,\lambda)$ the absorption cross-section. 
We assume that only silicate grains of sizes within $a_{\mathrm{alig,cold,min}}$ and $a_{\mathrm{alig,cold,max}}$ contribute to the polarized emission in the cold phase. Therefore, the polarized intensity of this phase is given by
\begin{equation}
    I_{\mathrm{pol,cold}} = f_\mathrm{sil} \times \int^{a_{\mathrm{alig,cold,max}}}_{a_\mathrm{alig,cold,min}} B_\lambda(T_{\mathrm{d,}i}(a)) \times\pi a^2\times Q_{\mathrm{abs,}i}(a,\lambda)\times a^{\alpha}\mathrm{d}a.
\end{equation}
Similarly, we model the total and polarized intensities from the hot phase as
\begin{equation}
    I_{\mathrm{hot}} = f_{\mathrm{scale}}\times\sum_i f_i \times \int^{a_{\mathrm{max}}}_{a_\mathrm{min}} B_\lambda(T_{\mathrm{d,}i}^{\prime}(a)) \times\pi a^2\times Q_{\mathrm{abs,}i}(a,\lambda)\times a^{\alpha}\mathrm{d}a
\end{equation}
and
\begin{equation}
    I_{\mathrm{pol,hot}} = f_{\mathrm{scale}} \times f_\mathrm{sil} \times \int^{a_{\mathrm{alig,hot,max}}}_{a_\mathrm{alig,hot,min}} B_\lambda(T_{\mathrm{d,}i}^{\prime}(a)) \times\pi a^2\times Q_{\mathrm{abs,}i}(a,\lambda)\times a^{\alpha}\mathrm{d}a,
\end{equation}
where we include a factor $f_{\mathrm{scale}}$ that determines the relative abundance of the material, i.e., gas and dust, in the hot phase in relation to that in the cold phase \citep{seifried_origin_2023}. Here, $T_{\mathrm{d,}i}^{\prime}(a)$ refers to either the un-attenuated temperature distribution $T_{\tau(1\,\mu\mathrm{m})=0.0}(a)$ or its mass-weighted average $T_{\mathrm{const},\tau(1\,\mu\mathrm{m})=0.0}$ (see Fig.~\ref{AttenuationTempFig}). \\
The wavelength-dependent polarization fraction then becomes
\begin{equation}
    p(\lambda) = \frac{I_{\mathrm{pol,cold}}+I_{\mathrm{pol,hot}}}{I_{\mathrm{cold}}+I_{\mathrm{hot}}}.
\end{equation}

\begin{figure*}
  \centering
  \includegraphics[width=0.96\textwidth]{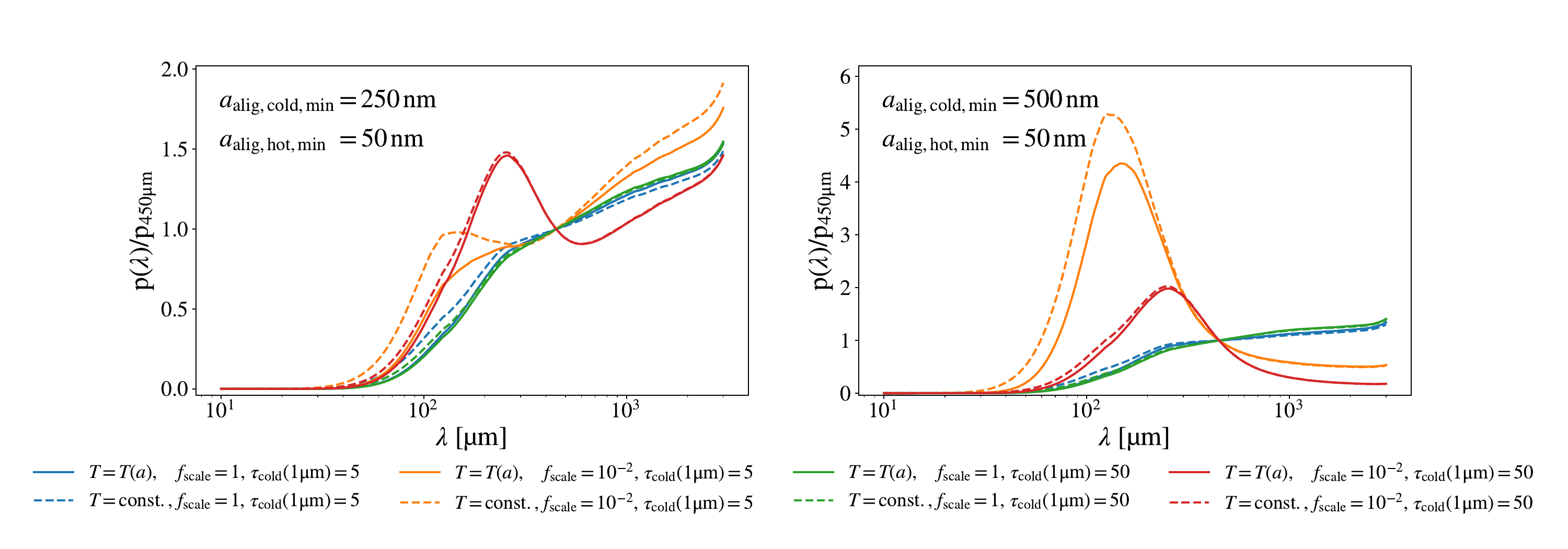}
  \caption{Semi-analytical polarization spectra for varying levels of relative abundance and attenuation between the hot and the cold phases. Exemplary polarization spectra, where the cold phase retains moderate levels of alignment (left panel), and where the alignment in the cold phase is lost (right panel).}
  \label{FulldustSpecFig}
\end{figure*}

\noindent For all cases considered, we assume that the upper size limit for alignment is higher than the maximum grain size, that is, $a_{\mathrm{alig,hot,max}} = a_{\mathrm{alig,cold,max}} = a_{\mathrm{max}}$ and that the grains in the hot phase have a lower size limit for alignment of $a_{\mathrm{alig,hot,min}}=50\,\mathrm{nm}$, which is similar to values found for dust in the diffuse ISM \citep{andersson_interstellar_2015}. For the cold phase, we consider a case where alignment is moderately well retained, setting $a_{\mathrm{alig,cold,min}}=250\,\mathrm{nm}$ and another case where alignment is lost, resulting in $a_{\mathrm{alig,cold,min}} = a_{\mathrm{max}} = 500\,\mathrm{nm}$. In Fig.~\ref{FulldustSpecFig} we compare the polarization spectrum obtained with the grain-size-dependent temperatures from Fig.~\ref{AttenuationTempFig} with the spectra obtained through their mass-weighted average temperatures at moderate levels of attenuation ($\tau_\mathrm{cold}(1\,\mu\mathrm{m)=5}$), as well as high levels of attenuation ($\tau_\mathrm{cold}(1\,\mu\mathrm{m)=50}$). Furthermore, we examine how the variation of $f_\mathrm{scale}$ influences this comparison, by considering $f_\mathrm{scale} = 1$ and $f_\mathrm{scale}=10^{-2}$. In the case where alignment is still retained (Fig.~\ref{FulldustSpecFig}, left), we generally see increasing or V-shaped spectra. Compared to the case where alignment in the colder regions is lost (Fig.~\ref{FulldustSpecFig}, right), we observe that the spectra that were previously V-shaped are now continuously decreasing, even at wavelengths $\lambda > 1\,\mathrm{mm}$. This shows that at longer wavelengths, the shape of the spectrum is defined by the contribution of the aligned population in the cold phase. In general, the difference between spectra where we consider grain-size-dependent temperatures and those where constant temperatures are applied is negligible compared to other uncertainties, such as differences in dust properties, temperatures, or the direction of the magnetic field lines with respect to the observer. The highest deviations are observed for the cases where $f_{\mathrm{scale}} = 10^{-2}$ and $\tau_{\mathrm{cold}}(1\,\mu\mathrm{m})=5$, regardless of whether the alignment is retained or not.

\section{The effect of normalization on the polarization spectrum slope}\label{AppC}
\begin{figure}[htbp]
  \centering
  \includegraphics[width=\columnwidth]{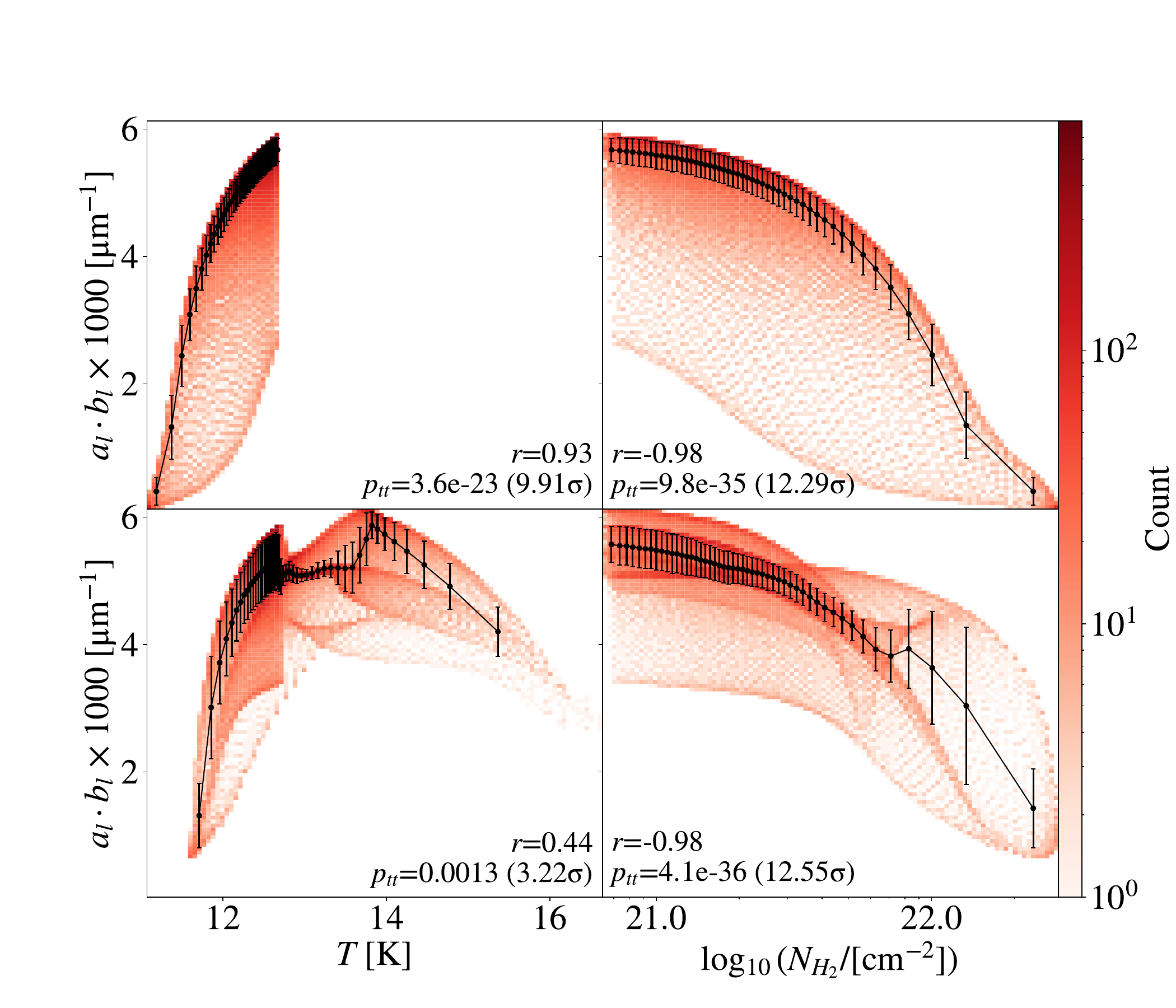}
  \caption{Same as Fig.~\ref{BlvsTNHFig}, but with the true slope $a_lb_l$ and without the use of normalization.}
  \label{BlvsTNHFig2}
\end{figure}
\begin{figure}[htbp]
  \centering
  \includegraphics[width=0.75\columnwidth]{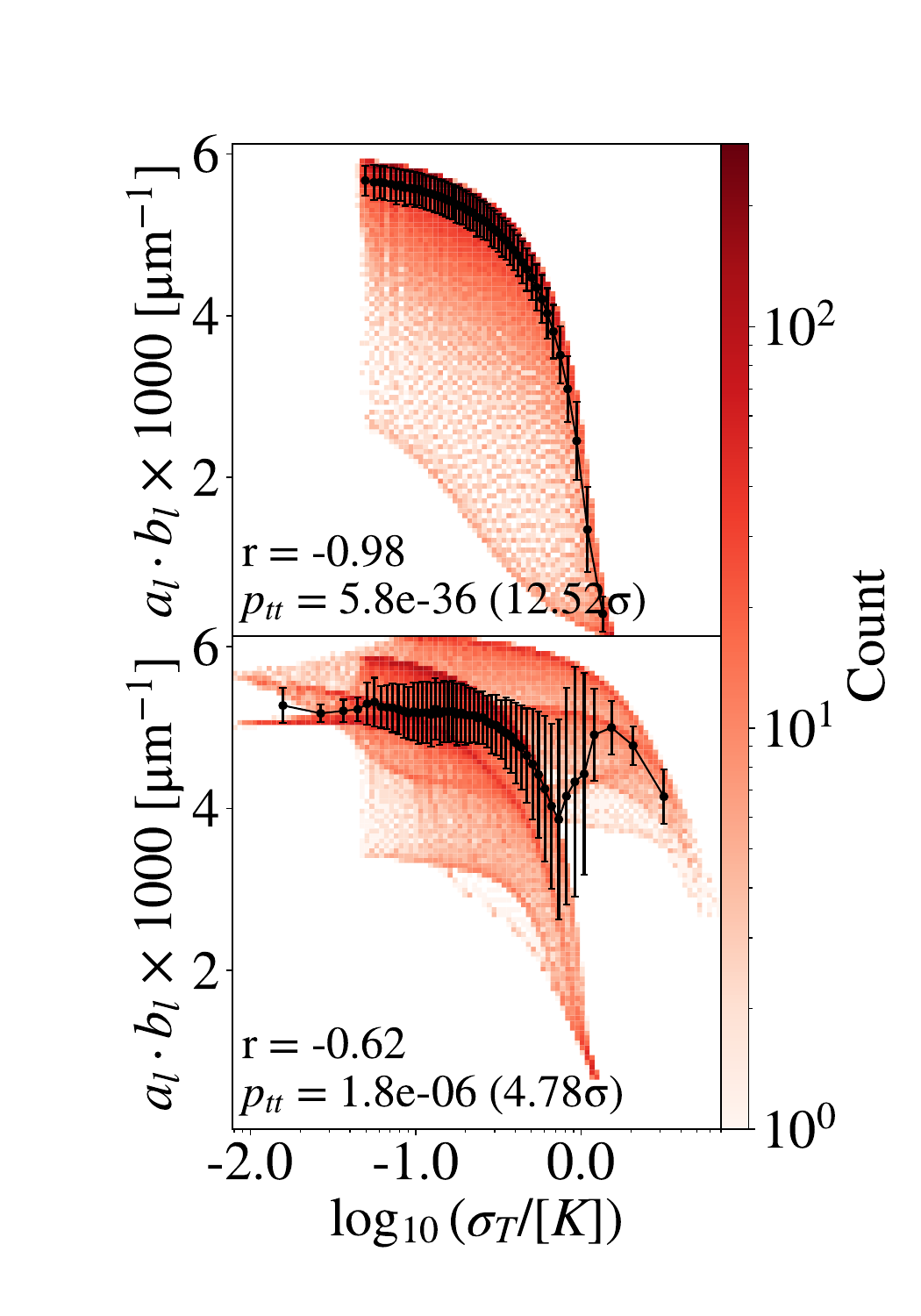}
  \caption{Same as Fig.~\ref{BlvsSigmaTFig}, but with the true slope $a_lb_l$ and without the use of normalization.}
  \label{BlvsSigmaTFig2}
\end{figure}

To investigate whether the initial increase in the slope of the polarization spectrum $b_l$, seen in Figs.~\ref{BlvsTNHFig} and~\ref{BlvsSigmaTFig}, can be attributed solely to the use of normalization, we fit Eq.~\ref{FitEq} to the same data we previously analyzed in Sect.~\ref{PolarisKap}.\\
Here, we cannot apply the assumption of $a_l\approx1$ anymore and refer to the product $a_l b_l$ as the slope instead. In Fig.~\ref{BlvsTNHFig2}, the Pearson $r$ correlation coefficient between the slope of the spectrum and the LOS temperature (left), as well as the column density (right), is calculated for the two setups, similar to Sect.~\ref{PolarisResultsKap}. The initial increase of the spectrum slope for more heterogeneous regions is no longer observed, resulting in a strong correlation with cloud properties. Fig.~\ref{BlvsSigmaTFig2} shows the level of correlation between the spectral slope and the standard deviation of the LOS temperature distribution, similar to Fig.~\ref{BlvsSigmaTFig}. For Setup one, where the ISRF acts as the only radiation source (top panel), we again see a strong correlation between these quantities. In the case where we consider a stellar source as well, we find a weaker correlation. This behavior arises even though the presence of either radiation source causes the spectrum slope to decrease. When calculating the correlation coefficient through the medians of chunked data points, differences in the sensitivity to $\sigma_T$ can lead to the median value increasing with higher levels of heterogeneity.\\
We note that an increase in the spectral slope with column density can alternatively arise through effects related to the optical depth. As \citet{lee_modeling_2024} mention, if the optical depth is high, polarized emission from within the cloud is partially absorbed before exiting, resulting in a reduction of the degree of polarization at shorter wavelengths and, in turn, a more positive slope of the spectrum. Since in our case $\tau\ll1$ holds for all wavelengths considered, this effect is negligible.

\end{document}